\documentclass[english,10pt]{IEEEtran}

\usepackage{footnote}

\usepackage{perpage}

\makeatletter

\def\blfootnote{\xdef\@thefnmark{}\@footnotetext}

\makeatother

\usepackage{amsthm}

\usepackage{amsmath}

\usepackage{amssymb}

\usepackage{mathtools}

\usepackage{graphicx}

\usepackage{babel}

\theoremstyle{plain}

\newtheorem{thm}{Theorem}

\newtheorem{lemma}{Lemma}

\newtheorem{corr}{Corollary}

\theoremstyle{definition}

\newtheorem*{defn}{Definition}

\newcommand{\bv}{\mathbf{v}}

\newcommand{\bx}{\mathbf{x}}

\newcommand{\bu}{\mathbf{u}}

\newcommand{\bt}{\mathbf{t}}

\newcommand{\bd}{\mathbf{d}}

\newcommand{\bc}{\mathbf{c}}

\newcommand{\be}{\mathbf{e}}

\newcommand{\bs}{\mathbf{s}}

\newcommand{\bp}{\mathbf{p}}

\newcommand{\dz}{\textrm{d}z}

\newcommand{\da}{\textrm{d}a}

\newcommand{\dx}{\textrm{d}x}

\newcommand{\dy}{\textrm{d}y}

\newcommand{\Umax}{U_{\max}}

\newcommand{\drho}{\textrm{d}\rho}

\newcommand{\Pmiss}{P_{\textrm{miss}}}

\newcommand{\Pmissnett}{P_{\textrm{miss}}^{\textrm{nett}}}

\newcommand{\Pmissone}{P_{\textrm{miss}}^{\textrm{one}}}

\newcommand{\deff}{d^{\textrm{eff}}}

\newcommand{\Eone}{E_{\textrm{all}}}

\newcommand{\separation}{\vspace{1.5pt}}



\begin{document}


\title{Scalable and Efficient Geographic Routing in Mobile Ad Hoc Wireless Networks}


\author{Dinesh~Ramasamy~and~Upamanyu~Madhow\\ University of California, Santa Barbara\\ Email:\{dineshr,~madhow\}@ece.ucsb.edu}

\maketitle 
\maketitle
\blfootnote{This work was supported by the Institute for Collaborative Biotechnologies through the grant W911NF-09-0001 from the U.S. Army Research Office. The content of the information does not necessarily reflect the position or the policy of the Government, and no official endorsement should be inferred. This paper was presented in part in \cite{isit_12}.}

\begin{abstract}

We propose and evaluate a scalable position-publish and an accompanying routing protocol which is efficient despite operating with imperfect information regarding the destination's location. The traffic generated by our position-publish protocol fits within the transport capacity of large mobile ad hoc networks (MANETs) with constant communication bandwidth allocated for routing overhead, even as the network size increases.  The routing protocol guarantees, with high probability, routes whose lengths are within a constant ``stretch'' factor of the shortest path from source to destination. The key idea underlying the scalability of the publish protocol is for each potential destination node to send location updates (with frequency decaying with distance) to a subset of network nodes, structured as annular regions around it (the natural approach of updating circular regions in distance-dependent fashion does not scale). The routing protocol must therefore account for the fact that the source and/or relay nodes may not have estimates of the destination's location (or may have stale estimates).  Spatial and temporal scaling of protocol parameters are chosen so as to guarantee scalability, route reliability and route stretch, and these analytical design prescriptions are verified using simulations.

\end{abstract}

\section{Introduction} \label{sec:intro}

Geographic routing is attractive for networks in which nodes know their own locations (e.g., using GPS) because a node only requires estimates of the locations of its immediate neighbors and of the destination node in order to forward a message. When the nodes in a network can move, a node can still maintain estimates of its neighbors' locations quite easily (the overhead for the local information exchanges for this purpose is small), but the bottleneck becomes global dissemination of information regarding the locations of moving destination nodes.  As observed in prior work (discussed in more detail shortly), this bottleneck can be alleviated by structuring location updates such that distant nodes get fewer updates, and live with a fuzzier view of the destination's location without excessively compromising route quality. This intuition is the starting point for the present paper, which provides an approach for {\it provably} scaling geographic routing to large mobile ad hoc networks (MANETs), while providing performance guarantees on route sub-optimality due to imperfect location information.


\begin{figure*}

\centering

\includegraphics[trim = 0.72in 0.4in 0in 0in, clip = true, width=0.625\columnwidth]{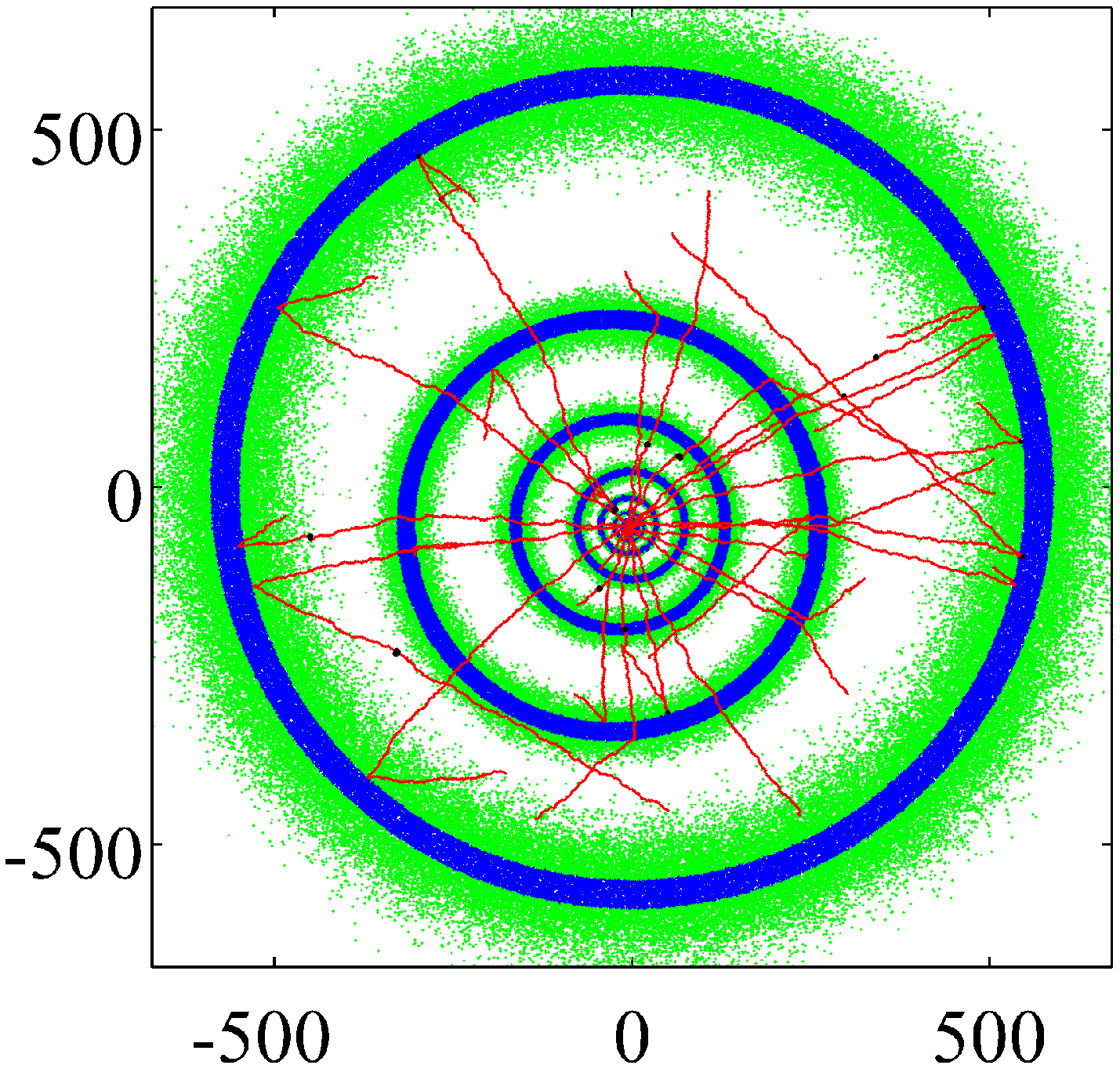}\quad\quad\quad\quad
\includegraphics[trim = 0.72in 0.4in 0in 0in, clip = true, width=0.625\columnwidth]{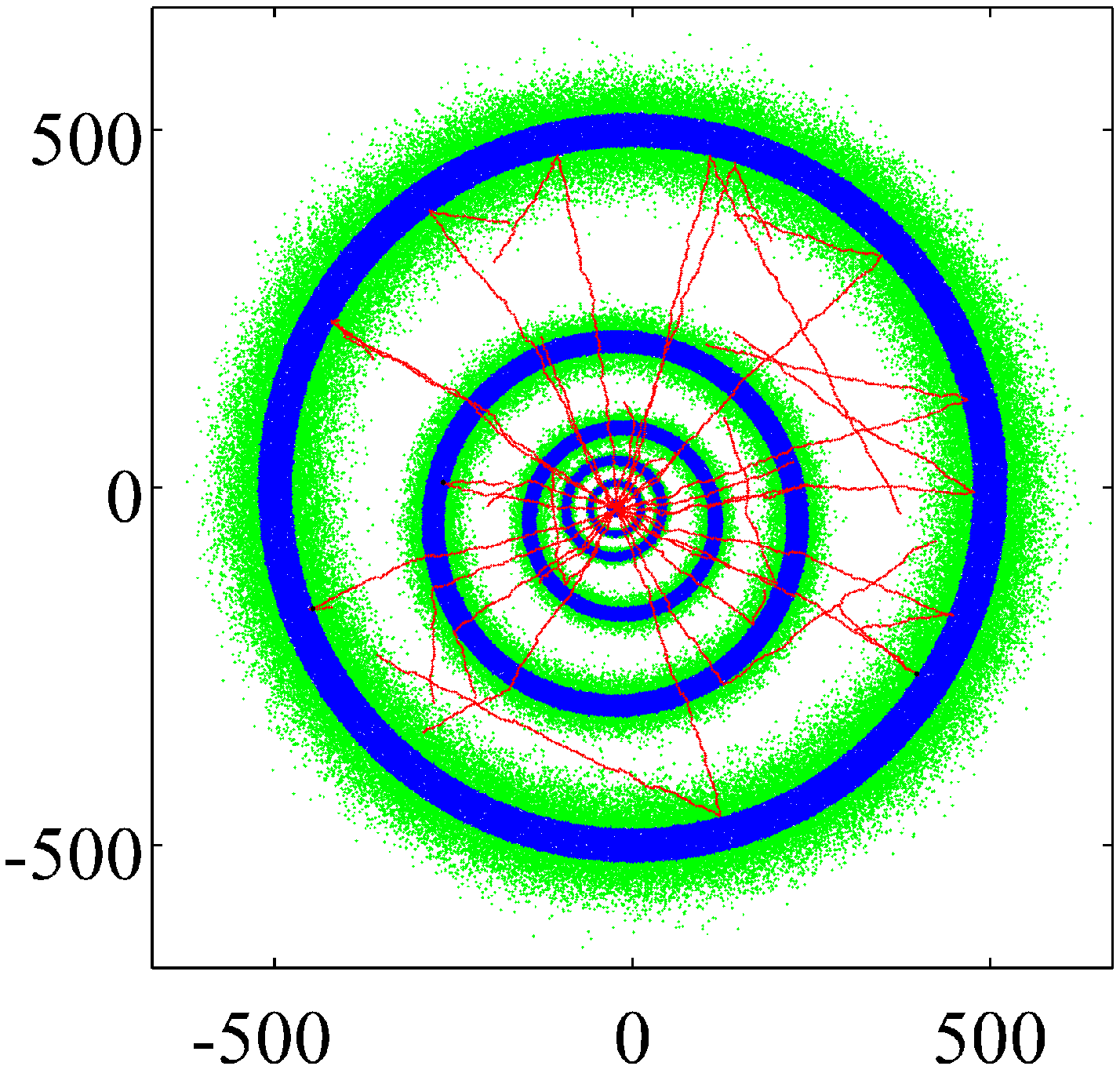}

\caption{ Overlaid routing trajectories (converging to a destination node) at a snapshot of the network for the proposed routing protocol for two different communication radii (smaller on the left). Blue dots indicate nodes with active position updates for the destination node; green dots indicate nodes with active position updates that however cannot be used because these nodes do not satisfy the spatial validity constraint of their position updates; red dots indicate relay nodes along packet trajectories using greedy geographic forwarding; black dots indicate greedy face traversal around voids \cite{Karp:GPSR}.}

\label{fig:trajectories}

\end{figure*}

For a routing protocol to be scalable, the traffic generated by routing updates must be within the network transport capacity obtained with a fixed communication bandwidth, bounds on which have been established in the pioneering work of Gupta and Kumar \cite{GuptaKumarWireless00}. In this paper, we provide a position-publish scheme which potential destination nodes use for global location updates, and show that (a) if all network nodes use this position-publish scheme, the resulting overhead is within the Gupta-Kumar bounds (can therefore be accommodated with constant bandwidth), and (b) information from the position-publish scheme enables routing with paths of length within a constant ``stretch'' factor of the shortest path from source to destination. Our contributions are summarized as follows:

\noindent $\bullet$ In order to characterize the stretch for greedy geographic forwarding, we derive necessary conditions on the communication radius to ensure that all greedy routing decisions  ``agree'' with the straight line joining the relay at which they are made to the position estimate used to make the decision. We precisely state this result in Theorem~\ref{thm:relaying:accuracy}.

\noindent $\bullet$ We show that, while bounded route stretch is feasible with uncertainty in destination location that scales with distance from the destination, the natural approach of simply reducing the frequency of location updates to distant nodes (which corresponds to updating {\it circular} regions), does not scale with network size. This implies that each potential destination must update a {\it subset} of network nodes regarding its locations.

\noindent $\bullet$ We show that scalability can be achieved by sending location updates to {\it annular} regions (the number of such regions scales as $\log n$, where $n$ is the number of nodes). Counting outward, the inner radius and thickness of these rings increase exponentially with their index. Key parameters of the location publish protocol are these exponents, as well as those of parameters determining the temporal validity of a location update and the ``quality'' of these location updates. We determine the constraints required on these parameters for achieving scalability and summarize them in Lemma~\ref{lemma:scalable}.

\noindent $\bullet$ Our routing protocol employs greedy geographic forwarding, with relay nodes overwriting information regarding the destination's location if they have ``better'' information than what is contained in the packet.  We determine constraints on the protocol parameters required to ensure that correct routing happens with high probability (reliable), and provide bounds on the worst case route stretch in terms of these parameters (Lemmas~\ref{lemma:reliable}~and~\ref{lemma:efficient}). We consolidate Lemmas~\ref{lemma:scalable},~\ref{lemma:reliable}~and~\ref{lemma:efficient} and present necessary conditions for scalable, reliable~\&~efficient (bounded stretch) routing in Theorem~\ref{thm:protocol:params}.

\noindent $\bullet$ We provide simulation results showing that choosing protocol parameters as prescribed by our analysis provides performance within the bounds guaranteed by our analysis.

The mobility model used for our analysis is two-dimensional (2D) Brownian motion. However, we note that our scalability results are broadly applicable to a large class of mobility models: scalability requires controlling the volume of traffic due to position updates to distant nodes, and when viewed from ``far enough'' away, a broad class of random mobility models ``look like'' Brownian motion.

The scalability of the position-publish scheme (while concurrently enabling routing with route stretch guarantees) is our primary interest. Therefore, we defer modifications to the position-publish protocol required to accommodate ``voids'' or ``holes'' in the deployment region to future work. We note however that voids can be handled for routing purposes using well-known techniques such as greedy face traversal \cite{Karp:GPSR}.

Before launching into a detailed exposition, we provide concrete insight into how the proposed protocol works via simulation results shown in Figure~\ref{fig:trajectories} (see caption for a brief overview, and Section \ref{sec:sim} for details).

\subsection{Related Work} \label{sec:related}

Since we are concerned with large-scale networks, our communication model and notion of scalability are guided by the relevant asymptotic results of Gupta and Kumar \cite{GuptaAsymptoticConn98} \cite{GuptaKumarWireless00}.  We postpone detailed discussion of these to Section \ref{sec:model}.

The literature on MANET routing and on geographic routing (for stationary or mobile nodes) is vast, hence we restrict attention here to prior work that is most closely related to our approach (many of the references we cite provide good discussions on the state of the art).  DREAM \cite{Basagni:DREAM1998} considers geographic routing when the frequency of location updates is reduced as the distance from the updating node increases.  While this intuition is the starting point for our scheme as well, we show that location updates made to all nodes as in DREAM are not scalable. A similar intuition is also behind the Hazy Sighted Link State (HSLS) algorithm in \cite{HazyStateLinkStateRoutingMobiHoc01}, in which link state updates are sent less frequently to distant nodes.  HSLS is designed based on minimization of the sum of the overhead due to route suboptimality and location updates.  However, the overhead computations in \cite{HazyStateLinkStateRoutingMobiHoc01} show that HSLS is not scalable. An intuitive reason for this is that all nodes must have a roughly consistent view of the network for successful link state routing, whereas geographic routing only requires that an appropriate subset of nodes have location updates from a given destination node. GLS \cite{Li:ScalableLocationService:Mobicom:00} is a spatially hierarchical quorum based scheme for position lookups, but is not designed to work in networks with pervasive movement.

MLS\cite{Flury:2006:MLS:EfficientAdHoc} proposes a ``lazy'' hierarchical position lookup service in which updates are published to certain fixed geographical regions. It is similar in spirit to our scheme, in that it is able to guarantee a constant route stretch without requiring that all nodes in the network obtain location updates, but the updates in our scheme are published to regions which are different, in general, for different nodes.  It is worth mentioning that MLS builds on an earlier scheme termed LLS\cite{LLS:Abraham:2004}, which structures location updates to areas centered around the destination node, as in our scheme.  The key difference of \cite{Flury:2006:MLS:EfficientAdHoc, LLS:Abraham:2004} from our work is they do not relate the routing overhead  to network transport capacity, and do not provide means to vary the tradeoff between route stretch and overhead.

Prior work \cite{GuptaAsymptoticConn98} derives the scaling of the communication radius needed for network connectivity. The critical radius needed for successful greedy geographic forwarding is derived in \cite{AsymptoticCriticalGreedyGeo:TC}. We derive necessary conditions on the communication radius to ensure that a routing decision made at a relay node using an arbitrary position estimate does not deviate ``too much'' from the straight line joining the relay to this estimate. Unlike \cite{AsymptoticCriticalGreedyGeo:TC}, our results hold even when the position estimates using which forwarding decisions are made do not correspond to current locations of network nodes. The node distribution in \cite{AsymptoticCriticalGreedyGeo:TC} is assumed to be given by a Poisson Point Process (PPP) of uniform density. While we make the assumption that the number of nodes is fixed and distributed uniformly and independently at random, we note that our scaling results also hold for the PPP model in \cite{AsymptoticCriticalGreedyGeo:TC}.

A preliminary version of this work appeared in our conference paper \cite{isit_12}. The present paper contains derivations related to routing reliability and efficiency omitted in \cite{isit_12}, as well as new results on the scaling of communication radius required to limit the worst-case disagreement between the actual and desired direction for forwarding.  The latter are required to ensure that our prescriptions for routing reliability and scalability are realizable.

For the routing scheme in this paper and the preceding references, mobility is a nuisance that increases routing overhead.  However, when delay in message delivery is not an issue, Grossglauser and Tse have shown in \cite{MobilityIncreasesCapacity:Grossglauser:2002} that mobility can actually help us get around the transport capacity limits derived by Gupta and Kumar \cite{GuptaKumarWireless00}. In a similar spirit, mobility can be exploited to reduce the overhead of location updates, as argued in \cite{Grossglauser:2006:EASE:EncounterHistory}\cite{Grossglauser:2003:AgeMatters:FRESH}.  However, this is not the regime of interest to us, since we are interested in delivering packets to their destinations with minimal delay.

\separation

\noindent \textbf{Outline:} We start in Section~\ref{sec:model} by describing the model and the accompanying scaling used in our computations. We focus on greedy geographic forwarding with location errors and give relevant definitions in Section~\ref{sec:geo_loc_errors_uncertainty}. We show in Section~\ref{sec:non_scalability} that the naive strategy of issuing updates (of necessary fidelity) to all nodes in the network does not scale. The proposed position-publish protocol, which overcomes this scalability bottleneck by issuing updates to a small subset of nodes is presented along with the accompanying routing protocol in Section~\ref{sec:proposed:protocol}. In Section~\ref{sec:derivation:protocol:parameter:choice}, we provide necessary conditions on the protocol parameters for scalable position-publish and reliable~\&~efficient routing. We report simulation results for one such choice in Section~\ref{sec:sim} before concluding with Section~\ref{sec:conclusions}.

\section{System Model} \label{sec:model}

We consider a network of $n$ nodes in the two-dimensional plane.  The deployment region is a square of area ${n}$ from which initial node positions are picked uniformly and independently at random. Therefore, the nominal node density is fixed at one node per unit area.

\separation
\noindent {\bf Connectivity:} We assume that the communication radius for all $n$ nodes is fixed at $r = r(n)$, and that it is chosen so that the network is connected. It is shown in \cite{GuptaAsymptoticConn98} that connectivity requires that $r$ must scale so that 
\begin{equation}
\pi r^{2} = \left(1+\epsilon\right)\log n,
\label{eq:comm_rad_choice}
\end{equation}
for any constant $\epsilon > 0$. While such a choice corresponds to a communication radius $r(n) = \Theta \left( \sqrt{\log n} \right)$, we note that we can scale down to a constant communication radius by scaling the deployment region as ${n}/{\log n}$ (along with suitably scaling other parameters) rather than as $n$.

\separation
\noindent {\bf Scalability:} We use the protocol model for interference proposed in \cite{GuptaKumarWireless00} for our scalability computations: Each transmission precludes the reception of any other transmission within a disc of radius $\left(1+\Delta\right)r$, where $r$ is the communication radius ($\Delta$ being an absolute constant). Thus, if the bandwidth available for communication is $W$, then the maximum number of simultaneous useful transmissions available per time slot, denoted by $T_A (n)$, scales as $\Theta\left(\left.{Wn}\middle/{ r^{2}(n)}\right.\right)$. Denoting by $T_U \left(n\right)$ the average of the total number of simultaneous transmissions needed per unit time to sustain a protocol (overhead) across all nodes, we employ the following definition for the scalability of a protocol.

\begin{defn}\label{defn:scalability}
We refer to a protocol as \textbf{scalable} if $T_U (n) = O\left(n\middle/r^2(n)\right)$. In this case, the overhead needed for the protocol can be accommodated with a suitably chosen constant bandwidth $W$. 
\end{defn}

The preceding definition assumes that the load induced on the network as a result of position updates is uniform in space and time (this does hold for our mobility model, which is described next).

\separation
\noindent {\bf Mobility Model:} Every node in the network is mobile, executing 2D Brownian motion of mean square velocity $2\sigma^2$, with reflection at the boundaries of the deployment region (assumed to be square for convenience). We note that for the choice of square deployment region with initial node positions picked independently and identically at random from the uniform distribution, 2D Brownian motion with reflection at the boundaries results in \emph{instantaneous} node positions (marginals in time) also given by the uniform distribution, with each node's position being independent of the other nodes in the network.

While we choose the Brownian motion model for its analytical tractability, we note that our scalability results hold more broadly: scalability depends on how distant nodes perceive the mobility of a destination node, and a large class of randomized models for \emph{local} mobility look like Brownian motion when viewed from far away and at large time scales. For example, consider a version of the random waypoint model\cite{Random:Waypoint:96} in which each node chooses a new speed $V_{l} \geq 0$ independently and identically from a distribution and direction $\Phi_{l}$ uniformly over $\left[0,2\pi\right]$ for a duration $D_l$, where the times $D_l > 0$ are independent and identically distributed random variables. It can be shown that, over large time scales, this model can be viewed as Brownian motion with mean square velocity $\left.{\left(\mathbb{E}V_1^2 ~ \mathbb{E}D_1^2\right)}\middle/{ \mathbb{E}D_1}\right.$.

\subsection{Greedy geographic forwarding with location errors}\label{sec:geo_loc_errors_uncertainty}

The routing protocol that we consider is the following: When a packet arrives at a node which is not the intended destination node, this node forwards the packet to the neighbor that is the closest to the current estimate of destination node's position (this position estimate may be available at the relay node or may have been appended to the packet by an earlier relay node). We refer to such a local routing strategy as \textbf{greedy geographic forwarding}. To facilitate this, we assume that every node has perfect knowledge of the location of its neighboring nodes.

We want to ensure that greedy geographic forwarding with imperfect location estimates is \emph{reliable} and that successful routes are \emph{efficient}. In the forthcoming discussions, we formally define these properties of greedy geographic forwarding protocols.

\begin{defn}\label{defn:reliable}
We refer to a routing protocol as \textbf{reliable} if it delivers packets to their destination with high probability (w.h.p.). We note that greedy geographic forwarding with \emph{perfect} location information is reliable when $\epsilon$ in \eqref{eq:comm_rad_choice} exceeds $\epsilon_0 \approx 1.6$ \cite{AsymptoticCriticalGreedyGeo:TC}.
\end{defn}

When information about the destination's location is imperfect, the natural approach is to route the packet along the best estimate of the direction of the destination, possibly updating this estimate after each hop, until we get close enough that the destination is within the communication radius. If the angle between the correct and the forwarded direction is $\theta$, then the progress towards the destination per unit distance traveled is $\cos \theta$, so that we would like $\theta$ to be small.

\begin{figure}

\centering

\includegraphics[clip = true, width=0.7\columnwidth]{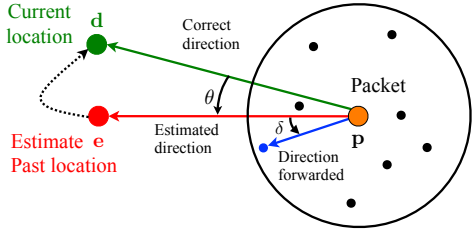}

\caption{Rate of progress depends on the angle between the forwarded and correct directions. The circle around the packet is the neighborhood of the relay node (given by the communication radius). }

\label{fig:Quasioptimal-routing}

\end{figure}

We now observe that we can afford to be sloppier in our estimate of the destination's location when we are farther away. Let us denote the distance between points $\mathbf{a}$ and $\mathbf{b}$ by $\ell\left(\mathbf{a},\mathbf{b}\right)$.

\begin{defn}\label{def:uncertainty}
We define the \textbf{uncertainty} $U$ of the position estimate $\be$ of the destination at $\bd$ available to a packet at $\bp$ as the ratio of distances $U =  \left.{\ell\left(\bd,\be\right)}\middle/{\ell\left(\bd,\bp\right)}\right.$.
\end{defn}

When we fix the uncertainty of the available estimate to $U$, it is easy to show that the worst case (largest) value of $\theta$ is given by $\sin \theta = U$. Thus, if we wish to ensure that the angle between the estimated direction and the correct direction is less than $\theta$, we can allow for the localization error to be larger when the packet is farther away (i.e., $\ell\left(\bd,\be\right)$ can be as large as $\ell\left(\bd,\bp\right)\times \sin\theta$).

\begin{defn}\label{def:stretch}
We define the ratio of the length of the source-destination packet trajectory to the source-destination distance to be the route \textbf{stretch}. 
\end{defn}


\begin{defn}\label{def:efficient}
The reciprocal of route stretch is a measure of routing efficiency and we refer to a routing protocol with a bounded stretch as an \textbf{efficient} protocol.
\end{defn}

\noindent {\it Bounded uncertainty leads to bounded stretch:} Now, suppose that the uncertainty seen by a packet is always less than $\Umax<1$, so that the worst case angle between the correct and estimated directions always satisfies $\theta \leq  \arcsin\left(\Umax\right)$.  This implies that $\cos \theta \geq \sqrt{1 - \Umax^2}$ and the route stretch will be bounded by $1/\sqrt{1 - \Umax^2}$.

The preceding argument assumes that the forwarded direction \emph{perfectly matches} the estimated direction. However, the neighbor of the relay node node (at $\bp$) which is the closest to the estimate $\be$ is never (with probability one, for our model of network nodes as points) on the line joining $\bp$ and $\be$. Therefore, we expect some disagreement between the actual direction along which a packet is forwarded and the {\it desired} direction corresponding to the estimate $\be$ used (we sketch this in Figure~\ref{fig:Quasioptimal-routing}). The amount of this disagreement depends on the availability of neighbors around the relay node along the estimated direction and thereabouts. Providing route stretch guarantees while taking this variability into account requires that we choose a large enough communication radius \eqref{eq:comm_rad_choice} by an appropriate choice of $\epsilon$. This ensures that w.h.p., the amount of disagreement between the forwarded direction and the desired direction (denoted by $\delta$ in Figure~\ref{fig:Quasioptimal-routing}) is small for all greedy forwarding decisions. We summarize this in the following theorem (proofs of both the theorem and its corollary are provided in Appendix~\ref{appendix:proof:lemma:relay:accuracy}). We note that this theorem holds for an \emph{arbitrary} position estimate, which need not correspond to the current location of any of the $n$ nodes in the network.

\begin{thm}
\label{thm:relaying:accuracy}
For any $0<\delta\leq\pi/3$, the following statement holds w.h.p. The maximum disagreement between the direction along which a packet is forwarded and the desired direction given by the straight line joining the relay with the estimate using which this greedy routing decision is being made is at most $\delta$, when $\epsilon$ in the choice of communication radius $r = \sqrt{\left.(1+\epsilon) \middle/\pi\right.\log n }$ is a large enough constant so that 
\begin{equation}
1 + \epsilon > \left.{\pi}\middle/{\left(\delta - \sin\delta\right)}\right.
\label{eq:slack:comm:rad:imperfect:estimates}
\end{equation}
and the estimate is at least $2r$ away from the relay node
\end{thm}

%

\begin{corr}
\label{corr:uncertainty:reliability}
Suppose that all nodes within $2\left(1-\Umax\right)^{-1} r$ of one another know each other's locations perfectly. When the uncertainty seen by packets is bounded by $\Umax<1$, routing with \emph{imperfect estimates} is \emph{reliable} if $\epsilon$ satisfies \eqref{eq:slack:comm:rad:imperfect:estimates} for some $0<\delta<\min\left\{\pi/3, \pi/2 - \arcsin\Umax\right\}$.
\end{corr}

\noindent\textit{Remarks:}

\noindent $\bullet$ It is possible to tighten the scaling of the communication radius needed for a particular choice of $\delta$ as prescribed by \eqref{eq:slack:comm:rad:imperfect:estimates} to 
$$
1+\epsilon > \left.{\pi}\middle/{\left(\delta - 0.5\sin(2\delta)\right)}\right..
$$
However, for simplicity of exposition, we only provide the proof of the looser scaling in \eqref{eq:slack:comm:rad:imperfect:estimates}.

 \noindent $\bullet$ When uncertainty is smaller than $\Umax$ and $\ell(\bp,\bd) > 2\left(1-\Umax\right)^{-1}r$, it can be shown that $\ell(\bp,\be) > 2r$. Therefore, from Theorem~\ref{thm:relaying:accuracy} we have that, if $\epsilon$ satisfies \eqref{eq:slack:comm:rad:imperfect:estimates}, the stretch of the segment of the trajectory from the packet source up until a distance of $2\left(1-\Umax\right)^{-1}r$ from the destination node is bounded by 
\begin{equation*}
\left.1\middle/\cos\left(\arcsin\left(\Umax\right) + \delta\right)\right.\leq \left(\sqrt{1-\Umax^2}-\sqrt{2}\delta\right)^{-1}\!\!\!\!.
\label{eq:stretch:bound:uncertainty}
\end{equation*}

A natural approach to guarantee a worst case route stretch is to employ a position-publish protocol that maintains uncertainty below a level $\Umax < 1$ throughout the network. We show in the next section that such protocols would not scale.  Before doing that, we round out this section by quantifying the cost of multicasting information to a specific region (which is a basic building block for the position-publish protocol discussed here).

\subsection{Cost of Multicast} \label{sec:multicast}

We note here for future use that the \emph{optimal} number of transmissions needed to multicast a message to all nodes in a connected region $A$, $C\left(A\right)=\Theta\left({\left|A\right|}/{r^{2}}\right)$.  

To see this, we note that, in order for every node in $A$ to have listened to the message at least once, the area $|A|$ has to be ``tiled'' by circles of area $\pi r^2$. Thus, $C(A) =\Omega \left(|A|\middle/r^2\right)$.

To provide an upper bound, we need a constructive scheme that multicasts messages to all nodes in $A$. When $\pi r^{2}=\left(1+\epsilon\right)\log n$, with $\epsilon$ a sufficiently large constant, we can use a result from \cite{GuptaKumarWireless00} to tile the area {\it a priori} into $\Theta\left({\left|A\right|}/{r^{2}}\right)$ tiles such that there exists at least one node per tile and every node in a tile can communicate with every other node in its tile and all the nodes in its neighboring tiles. Thus, the resultant network of tiles is connected. We designate one node per tile (say the one with the smallest node ID) to transmit and listen while others merely listen.  The first instant when the designated node in each tile receives the multicast message, it airs the message to all nodes in its range in a collision free manner (using an appropriate MAC) by means of a wireless broadcast.  Thus, the message is multicast to all nodes in $A$ with $\Theta\left({\left|A\right|}/{r^{2}}\right)$ transmissions, which proves that $C\left(A\right)=\Theta\left({\left|A\right|}/{r^{2}}\right)$.

Note that, even though nodes are mobile, the tiling of the network can be done {\it a priori} as in \cite{GuptaKumarWireless00}, and a node leader elected based on the node of smallest ID occupying the tile (this only requires nodes to have information regarding their neighbors, which any position-publish protocol provides).

We have been able to use the results in \cite{GuptaKumarWireless00} because for the initial node deployment and mobility model considered, the instantaneous distribution of nodes in the network is given by the uniform distribution, with each node's position being independent of the other nodes in the network

\section{A Non-Scalability Result}\label{sec:non_scalability}

While maintaining the uncertainty guarantees an upper bound on route stretch, we now show that maintaining a uniform uncertainty throughout the network, which requires updating all nodes in the network, does not scale.

In order to maintain uncertainty of at most $\Umax$, location updates from a particular node (say $v$) must reach all nodes that are a distance $z$ away from it if it moves a distance roughly equal to $\left(\Umax z\right) /\left(1+\Umax\right)$. For our Brownian motion model, the mean time to move this distance is $\left.{\left(\Umax^2 z^{2}\right)}\middle/{\left(2\sigma^{2}\left(1+\Umax\right)^2\right)}\right.$, and the average frequency of updates to these nodes is the reciprocal of this time. The area of a small ring at distance $z$ is $2\pi z~\dz$ and, as shown in Section \ref{sec:multicast}, the minimum number of transmissions needed to inform all nodes in this ring is $C\left(2\pi z~\dz\right)=\Theta\left({z~\dz}/{r^{2}}\right)$.

Remembering that the diameter of the network is $\Theta\left(\sqrt{n}\right)$, the average number of transmissions allocated to a node $v$ per unit time $t_{U}$ must satisfy:
\begin{align} 
t_{U} & \geq \frac{2\sigma^{2}C\left(\pi k_{1} r^{2}\right)}{k_{2}^{2}r^{2}}+ \int_{k_{1}r}^{k_{3}\sqrt{n}}\frac{2\sigma^{2}C\left(2\pi z~\dz\right)}{k_4 z^{2}}\label{eq:cost_update}\\ 
& =  \sigma^{2}\frac{k_5}{r^{2}}+\sigma^{2}\frac{k_6}{r^{2}}\log\left(\frac{k_{3}\sqrt{n}}{k_{1}r}\right)\nonumber 
\end{align} 
for some constants $k_{1}$, $k_{2}$, $k_{3}$, $k_{4}$, $k_5$ and $k_{6}$. The first term corresponds to broadcasts to a circle of radius bigger than $r$ to ensure all nodes have accurate lists of neighbors, while the second term corresponds to the location updates to distant nodes aimed at preserving uncertainty. The inequality in \eqref{eq:cost_update} is because we have ignored the rate needed to preserve updates in space (other network nodes are mobile and so updates made to a certain region in space will not be available in that region indefinitely). So $T_{U}\left(n\right)=n\times t_U = \Omega\left({\left(\sigma^{2}n\log n\right)}\middle/{r^{2}}\right)$.

\noindent {\it Maintaining uniform uncertainty does not scale:} The ratio of required overhead to sustainable capacity is therefore given by 
$$ 
\left.{T_{U} (n)}\middle/{T_A(n)}\right.=\Omega\left({\left(\sigma^{2}\log n\right)}\middle/{W}\right) 
$$ 
which blows up (albeit slowly) for large $n$. Thus, a strategy of maintaining an upper bound on uncertainty throughout the network (and thus bounding route stretch) does not scale.

Clearly, in order to provide guarantees on route stretch, the angle between the true and estimated directions towards the destination cannot be too large.  But what we have just shown implies that we must appropriately choose a {\it subset} of nodes to update in order to reduce the routing overhead enough that the protocol can scale.  This observation motivates the proposed protocol described in the next section.

\section{Proposed Protocol} \label{sec:proposed:protocol}

We now describe a \emph{scalable} position-publish protocol, and an accompanying greedy geographic forwarding protocol which works with imperfect position estimates and is \emph{reliable} and \emph{efficient}. Before presenting the details of the position-publish protocol and the routing protocol, we provide an overview of the position-publish protocol  and state the necessary conditions on the protocol parameters for scalable position-publish and for reliable~\&~efficient routing.

\subsection{Overview of the position-publish protocol}\label{sec:update-overview}

We give a summary of the position-publish protocol executed by a typical node that is a potential destination (we call this the {\it destination node} henceforth) while deferring the details to Section~\ref{sec:update}. The destination node directs its updates to geographic regions structured as annular rings around its current position, indexed as $i=0,1,...,K$. The position-publish algorithm is executed in a parallel fashion for each ring index.

 An update ring corresponding to index $i$ has inner radius of the $r_{i}$ and thickness $d_{i} \ll r_i$. Therefore, the geographical region to which an update is issued is specified by the center $\bc$ of the ring and its ring index $i$. An update issued to the $i$-th ring is retained for a duration of $T_i$ by the nodes that receive this update after which it is discarded. We refer to this time duration over which a particular update is retained as its \emph{lifetime} and those updates whose lifetimes have come to pass as \emph{expired} updates. The parameters $r_i,d_i$ and $T_i$, which define the ring index $i$ all scale exponentially with the ring index $i$.

\begin{figure}

\centering

\includegraphics[clip = true, width=0.4\columnwidth]{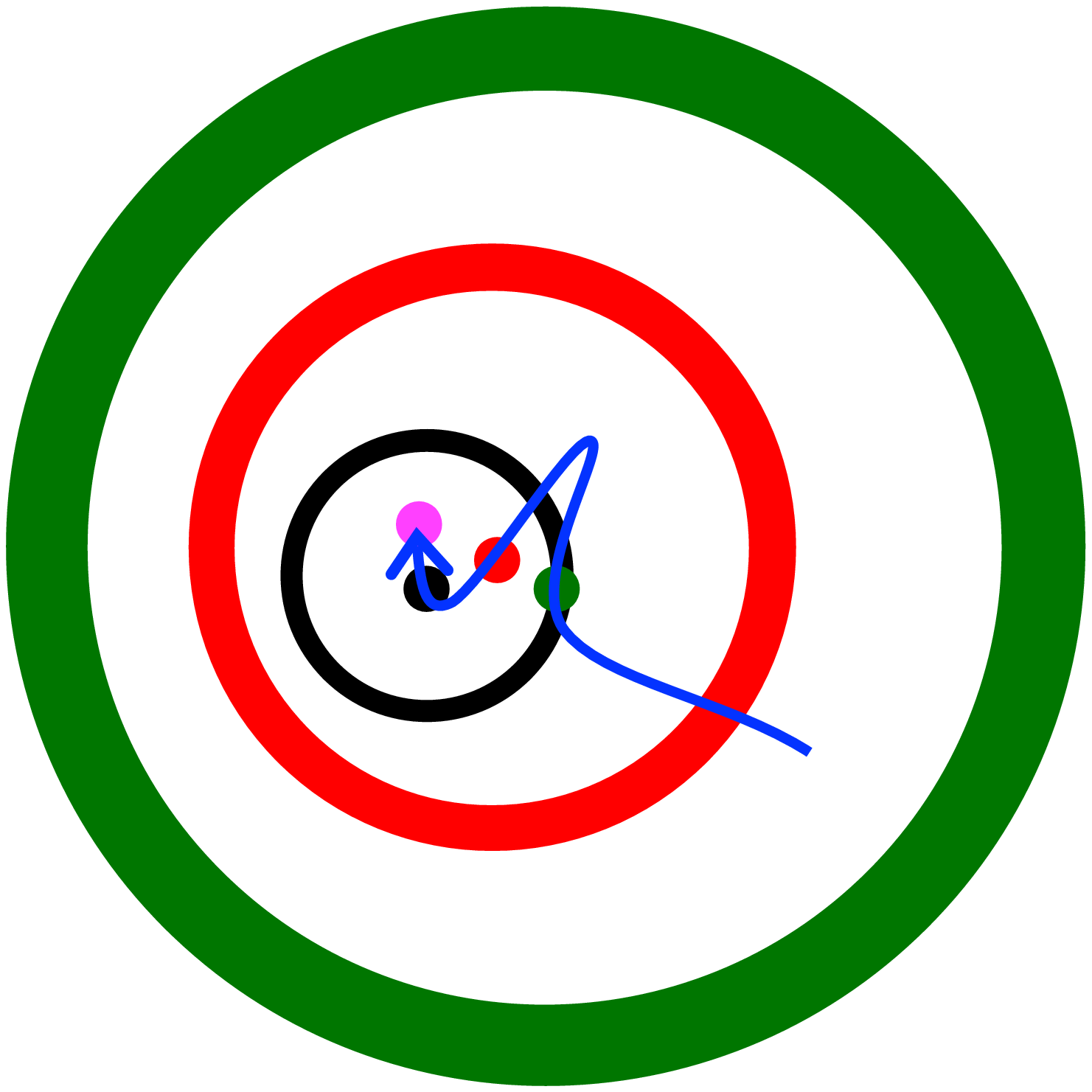}\quad\quad\includegraphics[clip = true, width=0.4\columnwidth]{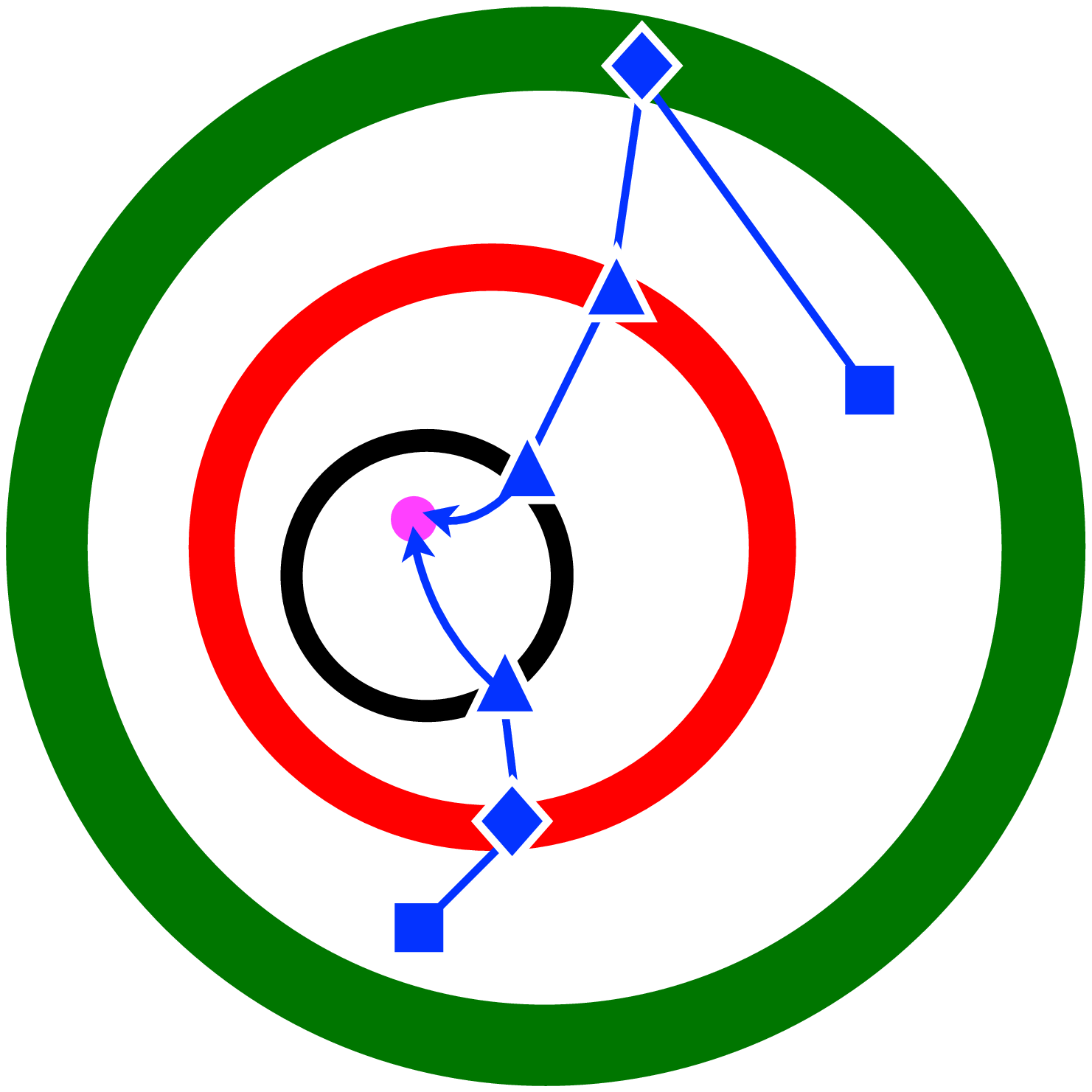}

\caption{Left: Update rings corresponding to three consecutive ring indices $l-1$ (black), $l$ (red) and $l+1$ (green). The position estimates (also centers of the update ring) are marked $\bullet$ on the destination's trajectory (blue) using corresponding colors. The current position of the destination is the magenta $\bullet$. Right: Two packet trajectories (blue) starting from nodes marked {\tiny $\blacksquare$} in between the $l$-th and $(l+1)$-th update rings converging to the destination (magenta $\bullet$). The packets are launched in arbitrary directions and acquire their first estimate (bootstrap) inside the $l$-th and $(l+1)$-th update rings respectively (marked $\blacklozenge$). They progressively refine their estimates when they cut through lower indexed rings (marked $\blacktriangle$).}

 \label{fig:rings}

\end{figure}

The position-publish protocol proactively publishes position updates to those ring indices whose updates are at the cusp of expiry, thereby ensuring that an update ring (which has not yet expired) corresponding to each of the $K+1$ update indices encircles the destination node. We depict the typical configuration of update rings around the destination node in Figure~\ref{fig:rings} (left). Three update rings corresponding to consecutive ring indices $\{l-1,l,l+1\}$ are highlighted in Figure~\ref{fig:rings}.  The position-publish algorithm runs in parallel for different ring indices (with different typical lifetimes $\{T_{l-1},T_l,T_{l+1}\}$). As a result, the three rings are centered around different points on the destination's trajectory.

\begin{figure}

\centering

\includegraphics[width=0.378\columnwidth]{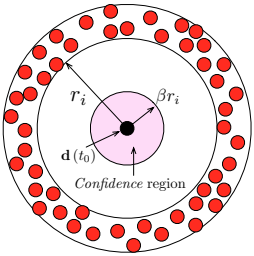}

\caption{Update made to the $i$-th ring at time $t_0$ by the destination node (in black). The nodes shaded red receive this update.}

 \label{fig:Publish_start}

\end{figure}

Denoting the position of the destination at time $t$ by $\bd(t)$, the position estimate $\be$ of an update points to the location of the destination node at the time of issue (i.e., an update made at time $t_0$ satisfies $\be = \bd\left(t_0\right)$). Each update also comes with a guarantee on the quality of its position estimate $\be $, which can be tuned by a parameter $\beta$ satisfying $0 < \beta < 1$: The destination node is understood to remain within a circle of radius $\beta r_i$  around this position estimate $\be$ (i.e., $\ell(\bd(t),\be) < \beta r_i$ until this update expires). We refer to this region as the \emph{confidence region} of the update. Figure~\ref{fig:Publish_start} illustrates a typical update ring and the {\it confidence region} associated with it. Confidence region guarantees are essential for the reliability and efficiency of the accompanying routing protocol. Two representative packet trajectories, which make use of the information disseminated by the position-publish protocol are sketched in Figure~\ref{fig:rings} (right).

\subsection{Protocol parameter choices} \label{sec:parameter_choices}

We summarize the regime of operation of the proposed protocol in Theorem~\ref{thm:protocol:params}. We provide a proof of Theorem~\ref{thm:protocol:params} in Section~\ref{sec:derivation:protocol:parameter:choice} via Lemmas~\ref{lemma:scalable},~\ref{lemma:reliable}~\&~\ref{lemma:efficient}.

The inner radius $r_i$, thickness $d_i$ and timer duration $T_i$ of update rings grow exponentially with the ring index $i$ as 
$$
r_i = r_0\alpha^{i},\quad d_i = d_0\alpha^{\mu i},\quad T_i = T_0\alpha^{\gamma i}.
$$
The zero-order ring defined by $r_0$, $d_0$ and $T_0$, the ring scaling exponents $\alpha$, $\mu $ and $\gamma $ and the confidence region parameter $\beta$ are the tunable parameters of the proposed position-publish protocol. Let $\Umax$ denote the maximum uncertainty seen by a packet after it acquires an initial estimate of the destination's position (we refer to this process of acquiring an initial estimate, as \emph{bootstrapping}, when we explain the routing protocol in Section~\ref{sec:routing}). We show in Appendix~\ref{sec:appendix:route:stretch} that $\Umax = \left.\alpha\beta\middle/(1-\beta)\right.$. We assume hereon that all nodes within $2\left(1-\Umax\right)^{-1}r$ of one another know each other's positions perfectly (via \emph{local} broadcasts), where $r$ is the communication radius, chosen to satisfy $\pi r^2 = (1+\epsilon)\log n$. Such local broadcasts are necessary for maintaining neighbor lists and can be accommodated within a constant bandwidth.

\begin{thm}\label{thm:protocol:params}
The proposed position-publish protocol is \emph{scalable} and the associated routing protocol is \emph{efficient} and \emph{reliable} when $r_0/\beta, d_0$ and $ \sqrt{T_0}/\sigma$ scale as $\Theta(r)$, $\alpha > 1$, $0 < \beta < \left.1\middle/(1+\alpha)\right.$, $1/3 < \mu < 1$, $1+ \mu < \gamma < \min\{2,4\mu\}$ and $ 1+ \epsilon > \left.\pi \middle/\left(\delta - \sin \delta\right)\right.$ for some constant $0<\delta<\min\left\{\pi/3, \pi/2 - \arcsin\Umax\right\}$.
\end{thm}

An example of parameter choices satisfying Theorem~\ref{thm:protocol:params}: (i) confidence region parameter $\beta=0.25$, (ii) order-zero ring specified by $r_{0}=r/\beta$, $d_{0} = 2r$, $T_0 =(1/8) \left(\beta r_0\middle/\sigma\right)^2$ and (iii) ring scaling parameters $\alpha=2$, $\mu=0.55$, $\gamma=1.95$. We perform simulations for this choice of parameters and present results in Section~\ref{sec:sim}.

\subsection{Position-publish protocol} \label{sec:update}

We now present the details of the position-publish protocol. There are two kinds of location updates: normal and abnormal updates.

\separation
\noindent{\bf Normal update:} A normal update published at time $t_0$ to ring $i$ (of radius $r_i$) specifies the center of the ring as the current location of the destination node. i.e., $\bc = {\bd} (t_0)$. The update points to the current location $\be= \bd (t_0)$ and has a lifetime of $T_{i}$ after which the nodes which receive the update discard it.

\separation
\noindent{\bf Abnormal update:} An abnormal update is sent when the destination leaves the confidence region for a prior \emph{normal} update before the timer for the latter update expires.  For example, for the normal update at time $t_0$ described above, if the destination node crosses the boundary of the confidence region at time $t_1 < t_0 + T_i$ (i.e., $\ell\left( \bd(t_1), \be\right) = \ell\left( \bd(t_1), \bd(t_0)\right) > \beta r_i$), then we send an abnormal update to the ring centered at the {\it prior} update. That is, we send an update specifying the current location $\be = \bd(t_1)$ to a ring of index $i$ centered at $\bc = \bd(t_0)$ with a timer $T_i - (t_1 - t_0)$ (spanning the remaining lifetime of the invalidated update).

When we choose the protocol parameters within the regime prescribed in Theorem~\ref{thm:protocol:params}, the probability of abnormal updates tends to zero as the ring index increases. However, we include abnormal updates to ensure that stretch guarantees are met.

\begin{figure}

\centering

\includegraphics[width=0.378\columnwidth]{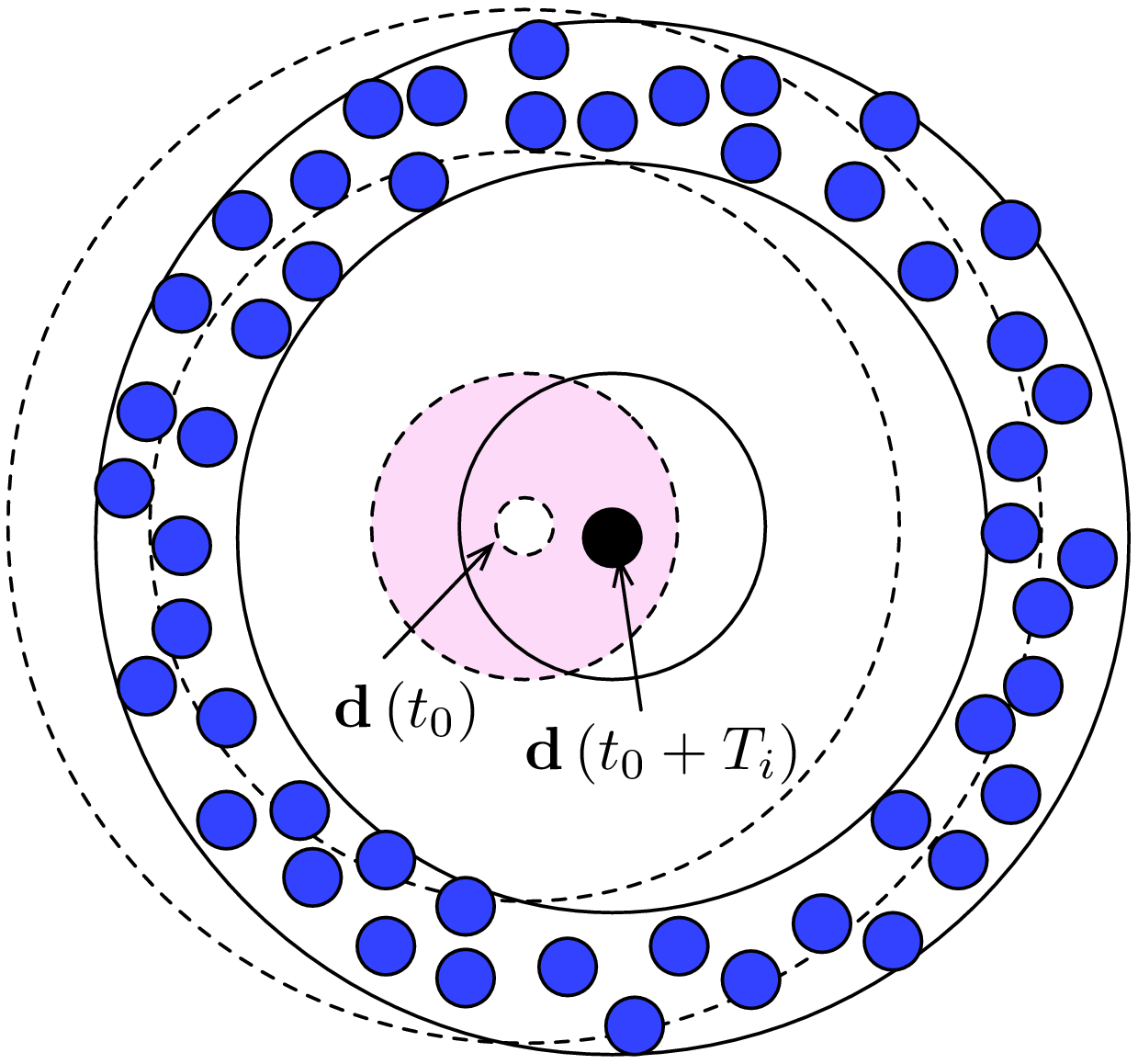}\space\space\includegraphics[width=0.44\columnwidth]{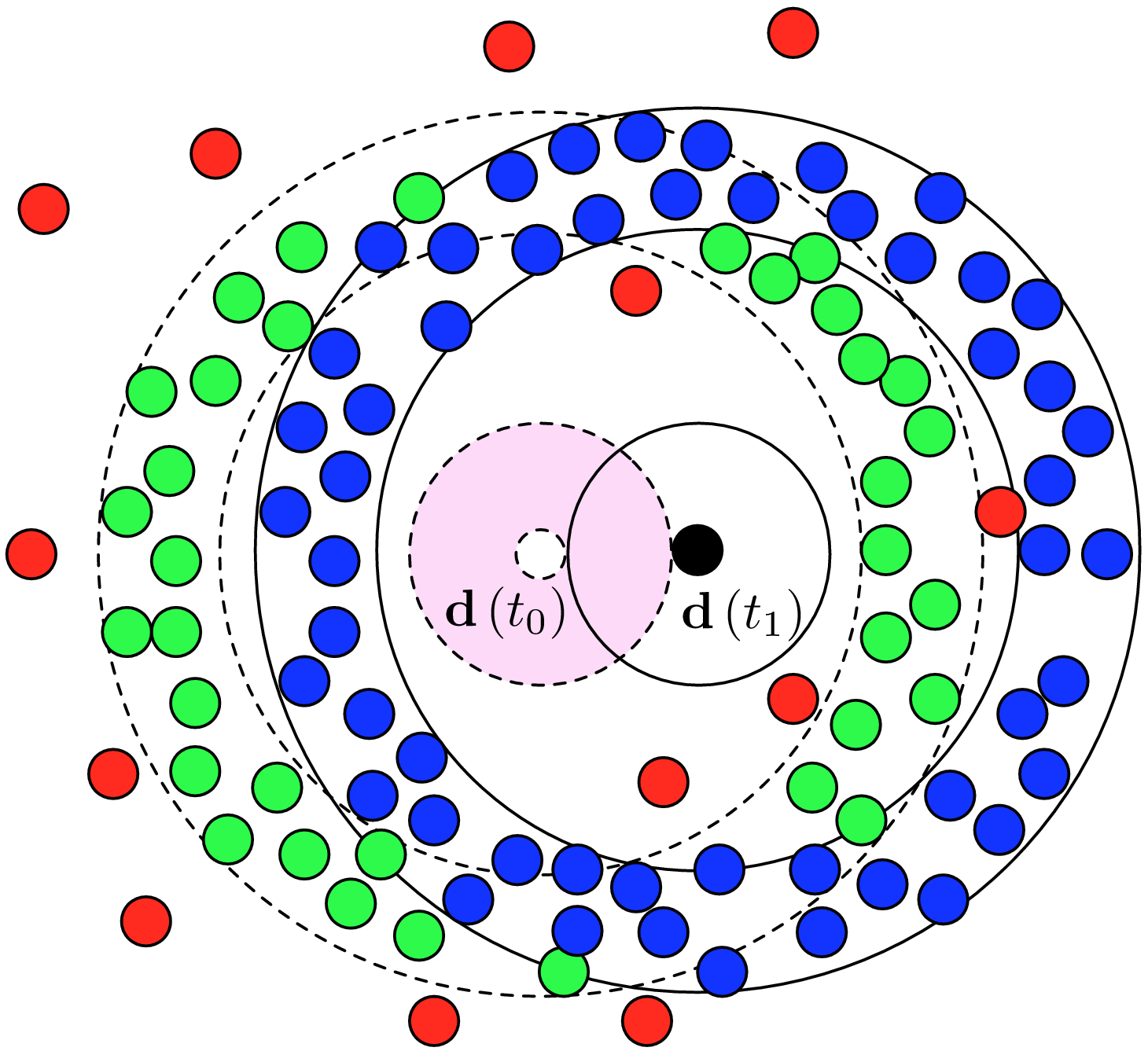} 

\caption{Left: Typical scenario of the destination node staying within the \textit{confidence region} of the update made at $t_0$, when it expires at $t_0+T_i$. A new \textit{normal} update of lifetime $T_i$ is made to the ring $i$ and is received by the blue relay nodes. Right: An unlikely situation where at time $t_1<t_0+T_i$, the destination node leaves the \textit{confidence region} of the update before it expires, thus requiring an \textit{abnormal} update (received by the green relays) of lifetime $T_i-\left(t_1-t_0\right)$ and a \textit{normal} update of lifetime $T_i$ (received by the blue relays). Relays marked red, \textit{outside} the two update rings possess stale unexpired updates made at time $t_0$ and these updates can be applied to packets only if these relays re-enter the ring centered at $\bd\left(t_0\right)$.}

\label{fig:Publish_normal_abnormal}

\end{figure}

\separation
 \noindent {\bf Triggers for new updates:} A normal update is performed whenever the timer for a prior normal update expires. This is depicted in Figure~\ref{fig:Publish_normal_abnormal} (left). When a destination node moves out of the confidence region of a \emph{normal} update whose timer has not expired, then \emph{two} updates are performed: a normal update to a ring centered around the current location and an abnormal update centered around the old location (at which the invalidated, but as yet unexpired, normal update was made).  The abnormal update lasts for the remaining lifetime of the invalidated update. This is shown in Figure~\ref{fig:Publish_normal_abnormal} (right). When a destination node moves out of the confidence region of an \emph{abnormal} update whose timer has not expired, then \emph{one} update is performed: an abnormal update centered around the old location which lasts for the remaining lifetime of the invalidated abnormal update. Abnormal updates prevent invalidated updates from influencing packet trajectories. The destination node maintains a list of updates published by it, so that it can publish new updates when these updates time out or when their guarantees are invalidated. Updates whose guarantees have been invalidated, are deleted from this list once the aforementioned compensatory action (of issuing new updates) is taken.

\separation
\noindent {\bf Spatial validity of updates:} An update (whose timer has not yet expired) can only be used for geographical forwarding if the relay node is in the ring to which the update was made (specified by its ring center $\bc$ and ring index $i$). Thus, once a node moves out of that ring, it can no longer use the information it received about the destination's location when in the ring. It will use this information if it moves into the update ring again. While this may seem overly restrictive, this constraint on the spatial validity of updates enables us to use abnormal updates to reinstate the confidence region guarantees needed for reliable~\&~efficient routing.

\separation
\noindent {\bf Update propagation:} In order to limit the traffic generated by a location update, the destination node sends the update packet in a specified direction until it hits the ring it is intended for, at which point it ``expands'' into a multicast message.  Specifically, the destination launches the packet in an arbitrarily chosen direction $\bu$, which is indicated in the packet.  Each intermediate node examines the packet to see if it is in the specified ring.  If not, it simply forwards the packet in the direction $\bu$.  Once the packet reaches a node in the update ring, that node repackages the update as a multicast packet for all nodes in the update ring. All nodes that receive this multicast message store the position update for the destination node (overwriting previous updates for the destination node with the most recent update). While large ``holes'' in the deployment region can disrupt update propagation and expansion, we note that this can be handled when we assume that the shape of the \emph{deployment region} (expected to be static) is known to all potential destination nodes. The destination nodes can thus choose launch direction(s) $\bu$ so as to avoid disruption of update propagation and expansion.

\subsection{Routing protocol} \label{sec:routing}

We now consider the problem of routing a packet to a destination which proactively publishes its location as described in Section~\ref{sec:update}. The packet contains a field indicating the destination identity, the ``best'' estimate of its location and the ring index \& time of update corresponding to this estimate. Intermediate nodes use this field for geographic forwarding, and are allowed to overwrite it if they have a ``better'' active estimate of the destination's location. An estimate is active only when the relay node's current location satisfies the spatial validity constraints of the update. An estimate is considered ``better'' only if its ring index is smaller than that of the packet or if its ring index is the same as that the packet, but the update is more recent than that of the packet. The ring index is given more importance than the time of update because of the guarantees given by the destination node through its layered update scheme.

If the source node does not possess an active update, then it chooses a random direction to relay the packet along: this is indicated in the packet by means of a vector indicating this direction (in the position estimate field), time of update $-\infty$ and ring index $\infty$.  Until the packet reaches a node with an active update all intermediate nodes relay the packet along this direction. When the packet hits a node with an active estimate, it is said to have \emph{bootstrapped}. If the packet reaches the boundary of the network before bootstrap, it bounces off the boundary by reflection (by a boundary node changing the direction field).

The parameters $\alpha$ and $\beta$ of the position-publish protocol limit the potential geometry of \emph{normal} update rings around the destination node and are chosen to ensure that before the packet reaches the estimate it possesses (say corresponding to the ring index $l+1$), it meets a smaller indexed update ring (the ring index $l$) and acquires the corresponding estimate (we detail this in Appendix~\ref{sec:appendix:route:stretch}). Therefore, the amount by which the packet's estimate $\be$ can disagree with the true location of the destination $\bd$, which can be no larger than the radius of the confidence region corresponding to the present estimate $\beta r_{l+1}$, progressively decreases after bootstrap and the packet eventually reaches the estimate corresponding to the ring indexed $0$, whose confidence region guarantee ensures that the destination is no further than $\beta r_0$ away from this location. This motivates the choice of the radius of the $0$-order ring, $r_0 = r/\beta$ in Section~\ref{sec:derivation:protocol:parameter:choice}, thereby ensuring successful packet delivery (we assume that nodes within $2\left(1-\Umax\right)^{-1}r > 2r > \beta r_0 = r$ know each other's locations perfectly). Two such converging packet trajectories are sketched in Figure~\ref{fig:rings} (right).

\section{Scalability, Reliability and Efficiency} \label{sec:derivation:protocol:parameter:choice}

We now derive the design guidelines for protocol parameter choices summarized in Theorem \ref{thm:protocol:params} in Section \ref{sec:parameter_choices}.  Recall that the inner radii of the update rings scale up exponentially with ring index: $r_{i}=r_{0}\alpha^{i}$, where $\alpha > 1$. So does the ring thickness, but at a slower rate: $d_{i}=d_{0}\alpha^{\mu i}$, $0<\mu<1$. The timer durations also scale up exponentially: $T_{i}=T_{0}\alpha^{\gamma i}$. The behavior of the protocol depends on the parameters: $r_0$, $d_0$, $T_0$, $\alpha$, $\beta$, $\gamma$ and $\mu$. In this section, we present three lemmas constraining the parameters for each of our design objectives: Lemma \ref{lemma:scalable} for scalability of the position-publish scheme, Lemma \ref{lemma:reliable} for routing reliability, and Lemma \ref{lemma:efficient} for routing efficiency.  Theorem \ref{thm:protocol:params} simply represents the intersection of the conditions for these three lemmas.

We start off by choosing the radius of the confidence region of the innermost update region, $\beta r_0$, equal to the communication radius $r$ (i.e., $r_0 = \Theta(\sqrt{\log n})$). This ensures that before the packet is forwarded to the node closest to the estimate corresponding to this zero-order ring, which can disagree with the true location of the destination by at most $\beta r_0 = r$, the packet is within a distance of $2r$ of the destination, thereby acquiring the true location (we assume nodes within $2\left(1-\Umax\right)^{-1}r > 2r$ of one another know each other's positions perfectly by means of local broadcasts). For this choice of $r_0$, the number of rings $K$ scales as $O\left(\log{n}\right)$. To see this, we note that we need the radius of the outermost update ring $r_K = r_0\alpha^{K}$ to roughly equal the network diameter $\sqrt{2n}$ (the deployment region is a square of area $n$), and this yields $K=\Theta\left(\log\left({n}\middle/{r_{0}^{2}}\right)\right)$.

\subsection{Position-publish Scalability} \label{sec:scalable}

Computing the average cost of updates to a particular ring index $i$ is the key step to computing the routing overhead.  For proving scalability, it is the behavior for large $i$ that is the most relevant.  For 2D Brownian motion, it can be shown that the probability of exiting a circle of radius $\beta r_{i}$ within the timer duration $T_i$ tends to zero, as long as $T_i$ grows slower than $r_i^2$. For $T_{i}=T_{0}\alpha^{\gamma i}$ and $r_{i} = r_{0} \alpha^{i}$, this is satisfied for large $i$ as long as $\gamma < 2$, which we henceforth take as a constraint. This implies that the rate of abnormal updates tends to zero (with ring index $i$), and that the update rate $F_i$ for ring $i$ is approximately ${1}/{T_i}$ for large $i$. In fact, $F_i \leq \Phi/T_i$ for all $i$, where $\Phi$ depends only on $\left.{\beta r_0}\middle/{\left(\sigma\sqrt{ T_0}\right)}\right.$. We choose $T_0$ so that $\left.r\middle/ {\left(\sigma\sqrt{ T_0}\right)}\right.=\Theta\left(1\right)$ and as a result $F_i = \Theta \left(1\middle/T_i \right)$ for all $i$.

As described in Section \ref{sec:update}, a position update to ring $i$ goes on a straight line until it hits the ring, and then is multicast in the ring. The area of the $i$-th ring ($i \geq 1$) is $A_i = \pi \left( (r_i+d_i)^2 - r_i^2 \right) = \pi d_i^2 + 2 \pi r_i d_i = \Theta \left( r_i d_i \right)$ (since the radius $r_i$ scales faster than the thickness $d_i$). From Section \ref{sec:multicast}, we know that the number of transmissions to multicast in this area is $C(A_i) = \Theta\left({\left|A_i\right|}\middle/{r^{2}}\right)$. Proceeding along the straight line takes $\Theta\left({r_i}\middle/{r}\right)$ transmissions, which can be ignored in comparison to the preceding.  Thus, the number of transmissions for an update to the $i$-th ring is $\nu_i = \Theta \left( {r_i d_i}\middle/{r^2} \right)$. The average rate of transmissions corresponding to updates for a typical destination node, which we term the {\it average overhead rate,} is therefore given by 
$$ 
t_U = \sum_{i=1}^K F_i \nu_i = \Theta \left( \sum_{i=1}^K \frac{r_i d_i}{r^2 T_i}  \right) 
$$ 
Plugging in the scaling for $r_i$, $d_i$ and $T_i$, we obtain that the average overhead rate is given by 
$$ 
t_U = \Theta \left( \frac{r_{0}d_{0}}{r^2 T_{0}} \sum_{i=1}^K \alpha^{\left(1+\mu - \gamma\right)i}\right)
$$ 
As $n$ gets large, so does the number of rings $K$, so that the preceding summation converges when $ \alpha^{1 + \mu - \gamma} < 1$. Since we need $\alpha > 1$ in order to exponentially expand the rings, we must have $1+ \mu - \gamma < 0$ as a necessary condition. As $n$ grows, we have already noted that $r_0$ scales as $\Theta(r/\beta)$ and ${T_0}$ as $\Theta\left(r^2\middle/\sigma^2\right)$. For this scaling, we show in Section~\ref{sec:reliability} that we need $d_0=\Theta(r)$ for reliable routing. We now have that 
\begin{equation*}
T_U = n\times t_U = \Theta \left({\sigma^2}\middle/{r^2} \right) 
\end{equation*}
which matches the throughput available per node $T_A$ for a fixed bandwidth.

\begin{lemma}\label{lemma:scalable}
The proposed position-publish protocol is \emph{scalable} when $1+ \mu < \gamma < 2$, $\beta r_0 = \Theta(r)$, $d_0=\Theta(r)$ and $T_0=\Theta\left(r^2\middle/\sigma^2\right)$.
\end{lemma}

\subsection{Routing Reliability} \label{sec:reliability}

We have analyzed the update protocol to determine conditions for scalability.  We now analyze the routing protocol to determine conditions that ensure reliable routing. After an update is made to nodes in a ring, some of these nodes may leave the ring. When a packet being routed to the destination hits the ring, therefore, the relay nodes it sees may be ones which moved in after the currently active update was made. According to the proposed routing protocol, when the packet meets such nodes which have estimates of the destination's location worse than its own (including not having any estimate of the destination's location), it simply continues in the direction it is going. Thus, in order for a packet to take advantage of an active update for ring $i$ once it hits it, it suffices that at least one of the nodes it meets as it is cutting through the ring has an active update corresponding to ring $i$. If this does not happen, we say that the packet has ``missed'' the $i$-th ring. The lifetime $T_i$ of normal updates must be short enough that the probability of a miss tends to zero, which imposes additional conditions on the protocol parameters, as we show here.

The worst case scenario for missing a ring is the following scenario: (i) The packet is relayed radially across it, since it meets fewer relay nodes along the ring, and hence a smaller probability of meeting a node with an active update (ii) The time since update issue is $T_i$ (just prior to update expiry). This is because the density of relay nodes with updates inside the update ring (the region where the update is spatially valid) decreases as time from update issue increases and the packet is least likely to meet a relay node with an active update just prior to update expiry. We consider this worst case scenario and use the following asymptotics (for outer rings; large $i$): $r_i/\sqrt{T_i} \rightarrow \infty$, $d_i/r_i \rightarrow 0$ and $\left.d_i\middle/\left(\sigma\sqrt{T_i}\right)\right. \rightarrow 0$ to give an upper bound on the miss probability $\Pmiss(i)$:
\begin{equation} 
\log \Pmiss(i)  \lessapprox  -\left.{d_{i}^{2}}\middle/{\left(\sigma r\sqrt{2\pi  T_{i}}\right)}\right..
\label{eq:pmiss_asymptotics} 
\end{equation}
We provide the details of this derivation in Appendix~\ref{sec:appendix:Pmiss}. Since $\sigma\sqrt{T_0} = \Theta(\beta r_0) = \Theta(r)$, if we ensure that $d_0$ grows at least as fast as $r$, we can bound the probability of missing the innermost ring ($i=0$) for all network sizes. Thus, we set $d_0 = \Theta(r)$.

We prove in Appendix~\ref{sec:appendix:route:stretch} that when no misses occur, the confidence region guarantees of active updates ensure that the uncertainty seen by a packet after bootstrap is bounded by $\Umax = \left.\alpha \beta \middle/(1 - \beta)\right.$. When we make the following assumptions: (i) $\beta$ is chosen so that $\Umax < 1$ (ii) all nodes broadcast their current positions to other nodes within $2\left(1-\Umax\right)^{-1}r$ (iii) $\epsilon$ in $\pi r^2 = (1+\epsilon) \log n$ is a large enough constant given by  \eqref{eq:slack:comm:rad:imperfect:estimates} for some $\delta < \min\left\{\pi/3,\arcsin\Umax\right\}$, we can invoke the corollary of Theorem~\ref{thm:relaying:accuracy} to guarantee routing reliability. Therefore, the overall probability of routing failure $\Pmissnett$ can be upper bounded by the probability of missing at least one ring $\Pmissone$ (so that the uncertainty can no longer be bounded by $\Umax$). Using the union bound, we have
\begin{equation}
\Pmissone \leq \sum_{i=0}^{i=K} \Pmiss(i) \lessapprox \sum_{i=0}^{i=K} \exp\left(-{d_{i}^{2}}\middle/\left({\sigma r\sqrt{2\pi T_{i}}}\right)\right).
\label{eq:pmiss_nett} 
\end{equation}
Since $\left.d_0^2\middle/\left(r\sigma\sqrt{T_0}\right)\right. = \Theta(1)$, we note that $\Pmissone$ is bounded for all network sizes (and thus, so is $\Pmissnett$) if $\gamma < 4\mu$. By choosing $\left.d_0^2\middle/\left(r\sigma\sqrt{T_0}\right)\right.$ to be large enough, we can drive the bound on $\Pmissone$ given by \eqref{eq:pmiss_nett} below any desired level.

\begin{lemma}\label{lemma:reliable}
The proposed routing protocol is \emph{reliable} when $\gamma < 4 \mu$, $0<\beta <  \left.1\middle/(1+\alpha)\right.$,  $d_0=\Theta(r)$, $T_0=\Theta\left(r^2\middle/\sigma^2\right)$ and $\epsilon$ in \eqref{eq:comm_rad_choice} is large enough so that \eqref{eq:slack:comm:rad:imperfect:estimates} holds for some $\delta<\min\left\{\pi/3, \pi/2 - \arcsin\Umax\right\}$.
\end{lemma}

\subsection{Routing Efficiency} \label{sec:stretch}

In order to bound the route stretch, we must account for the fact that, since location updates are sent to only a subset of nodes, the source node need not have an active update for the destination. In this case, the packet travels an additional distance in an arbitrarily chosen direction until it hits a node with an active update and we refer to this process of acquiring an initial estimate as the bootstrapping process. Our bound on route stretch must account for this additional distance. Once the packet does encounter a node with an active update, we use uncertainty, which we show is bounded by $\Umax = \left.\alpha\beta\middle/\left(1-\beta\right)\right.$  provided the packet does not miss smaller indexed rings thereafter, to bound the route stretch. The resulting bound on the route stretch is given by (we assume that $\epsilon$ in the choice of $r$ is large enough so that the direction along which a packet is forwarded closely matches that corresponding to the position update used; i.e., $\epsilon$ chosen to satisfy \eqref{eq:slack:comm:rad:imperfect:estimates} for small $\delta$):
\begin{equation}
\label{eq:stretch_expression} 
\sqrt{1+\left(\sqrt{\frac{ \alpha^{2}\left(1+\beta\right)^2}{\left(1-\beta\right)^2 }- 1} + \frac{\alpha\left(1+\beta\right)}{\sqrt{\left(1-\beta\right)^2 - \alpha^2\beta^2}}\right)^{2}} 
\end{equation}
Details of the derivation can be found in Appendix~\ref{sec:appendix:route:stretch}. We note that this bound is finite if $ \Umax =  \left.\alpha\beta\middle/\left(1-\beta\right)\right. < 1$.

\begin{lemma}\label{lemma:efficient}
Packet trajectories are within a constant \emph{stretch} factor of the source-destination distance (\emph{efficient}) when routing is \emph{reliable} (no additional constraints).
\end{lemma}

\section{Simulation results}\label{sec:sim}

We perform simulations of the position-publish and routing protocol for a particular destination node, for the following scenario: Number of nodes $n=1.8\times10^{6}$, node density $1$, mobility model being 2D Brownian motion with parameter $\sigma^{2}=1$ and the deployment area is a square of side $\sqrt{n}$. We choose the communication radius $r$ to be $\sqrt{\left(1+\epsilon\right)\log n/\pi}$ and report results for both $\epsilon=0$ and $\epsilon=2$. The parameters of the position-publish protocol are: (i) confidence region parameter $\beta=0.25$, (ii) order-zero ring specified by $r_{0}=r/\beta$, $d_{0} = 2r$, $T_0 =(1/8) \left(\beta r_0\middle/\sigma\right)^2$ and (iii) ring scaling parameters $\alpha=2$, $\mu=0.55$, $\gamma=1.95$.

To get a concrete sense of what these numbers mean, we choose the units of distance so that the communication radius is $r=100$m (for $\epsilon = 0$). The deployment area is $63$ km by $63$ km, with a node density of $458$ nodes per square km. Now we choose units of time so that RMS motion over one second $\sigma\sqrt{2}$, to be $10$m (consistent with vehicular speeds), we see that the lifetime of updates to the update rings of radii $3.2,6.4,12.8~\&~25.6$ km ($i = 3$ to $i=6$) and thickness $0.63,0.92,1.35~\&~1.97$ km are $0.40,1.55,5.98~\&~23.10$ hours respectively.

\begin{figure}

\centering

\includegraphics[clip =true, trim = 0.625in 0in 0in 0in, width=0.45\columnwidth]{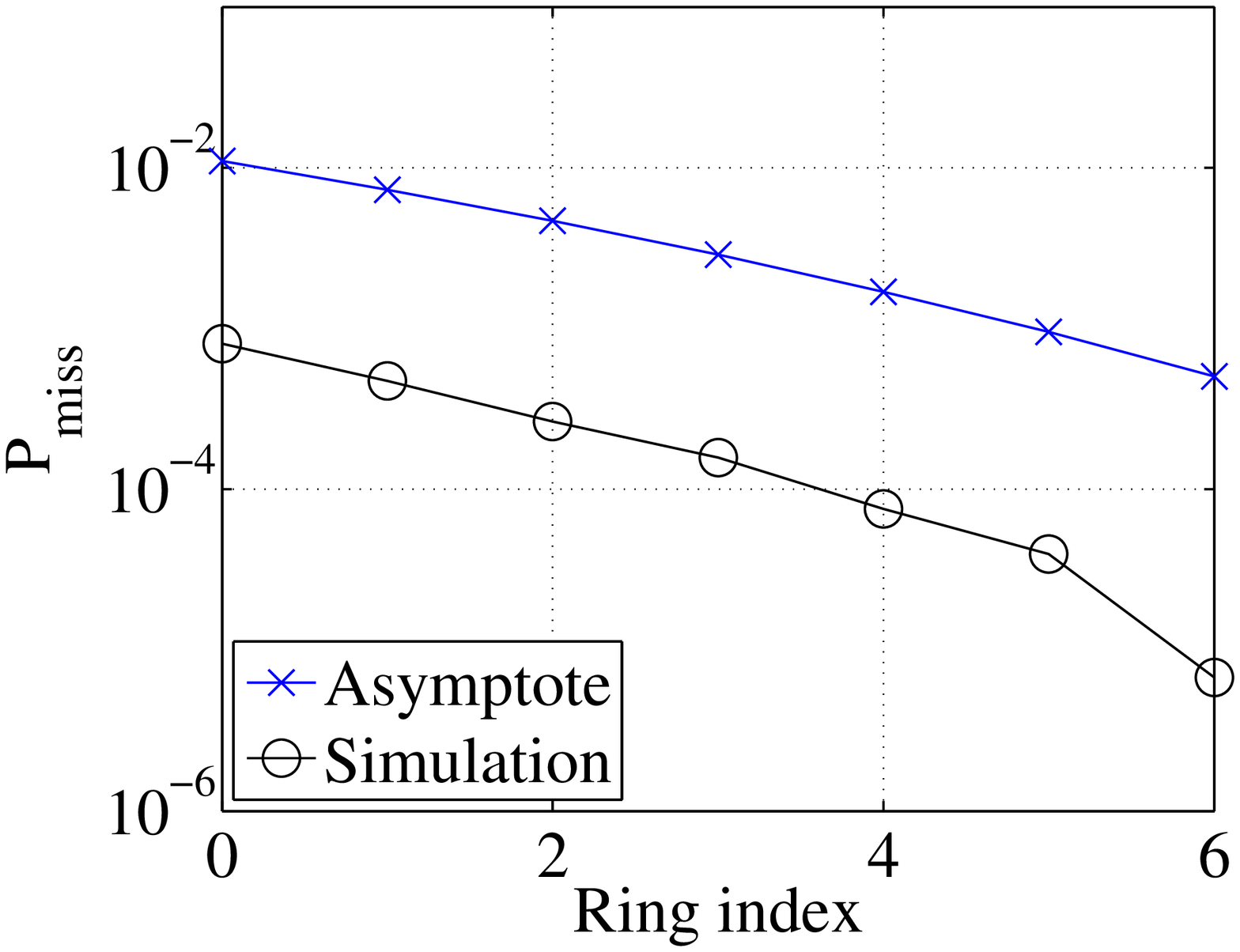}
~\includegraphics[clip =true, trim = 0.625in 0in 0in 0in, width=0.45\columnwidth]{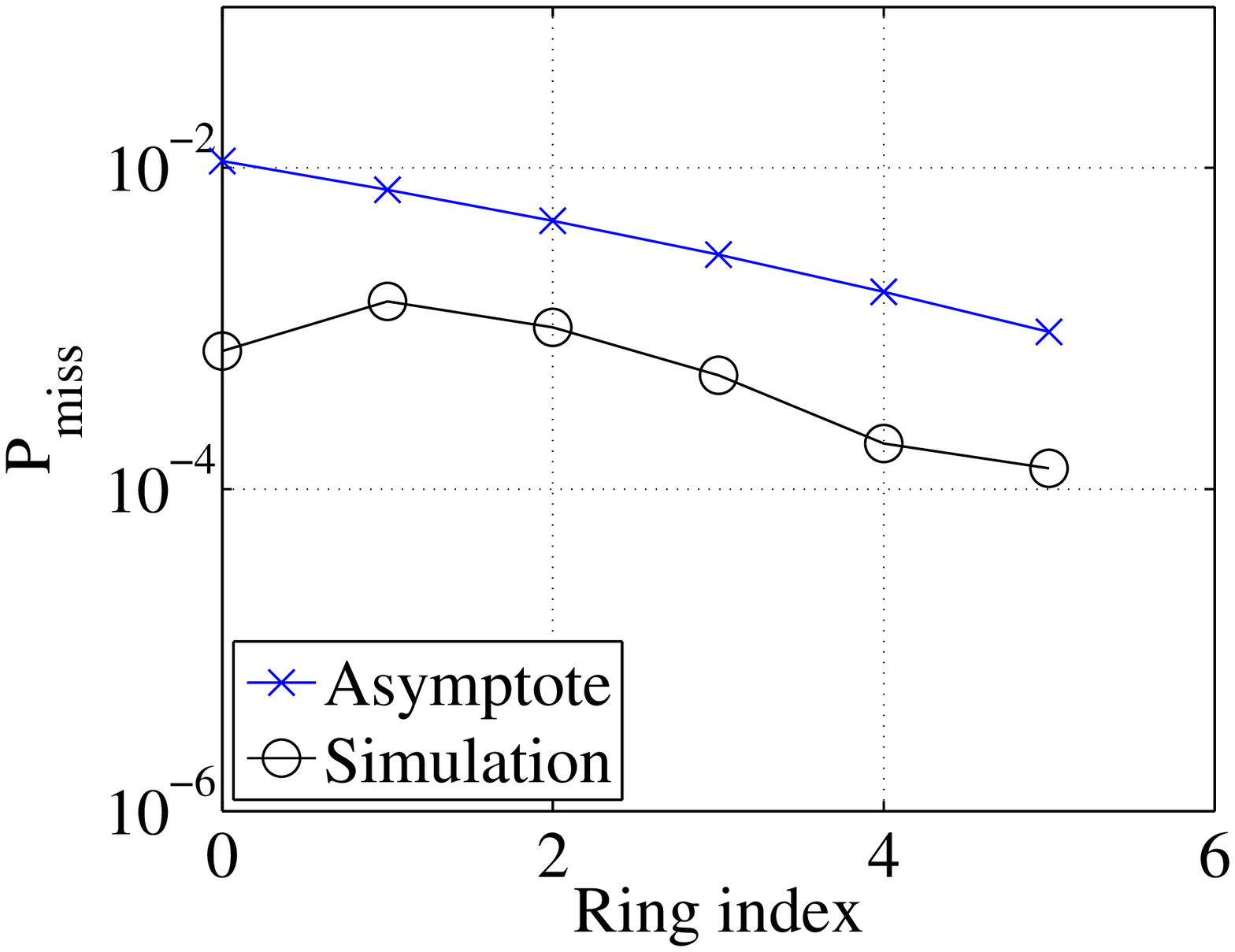}

\caption{Probability of missing a ring for $r$ corresponding to $\epsilon=0$ (left) and $\epsilon=2$ (right) for radial traversal just prior to update expiry (worst case)}

\label{fig:pmiss_asymp_empirical}

\end{figure}

\separation
\noindent {\bf Probability of a packet missing updated nodes in a ring:} We first compare the asymptotic estimate of the worst case miss probability in \eqref{eq:pmiss_asymptotics} with simulations of the worst case (radial trajectory \& just prior to update expiry) in Figure~\ref{fig:pmiss_asymp_empirical}. Note that the match is better for $\epsilon=2$, because the average number of relay nodes seen by a packet decreases with an increase in $\epsilon$ and approaches the lower bound of $d_{i}/r$ used in the derivation of the upper bound on $\Pmiss(i)$ (\eqref{eq:pmiss_min_hops} in Appendix~\ref{sec:appendix:Pmiss}).

\separation
\noindent {\bf Trajectories:} We now refer back to the simulated trajectories previewed in Figure~\ref{fig:trajectories} (Left: $\epsilon =0$, Right: $\epsilon = 2$) (we ignore edge effects by focusing only on trajectories that start inside the outermost update ring while noting that network boundaries can be handled either by using a specialized update ring, or via packet ``reflection'' at the boundaries).  We note that, for larger communication radius ($\epsilon =2$), the trajectories are straighter, as there are many nodes available in each direction around a relay node. For $\epsilon=0$, which is below the threshold \cite{AsymptoticCriticalGreedyGeo:TC} for asymptotic success of greedy geographic forwarding, trajectories hit voids frequently.  However, using the standard technique of greedy left hand traversal of voids \cite{Karp:GPSR} (these segments of the trajectory are marked in black), route failure rates are reduced to a small level.

\begin{figure}

\centering

 \includegraphics[clip =true, trim = 0in 0.39in 0in 0in, width=0.45\columnwidth]{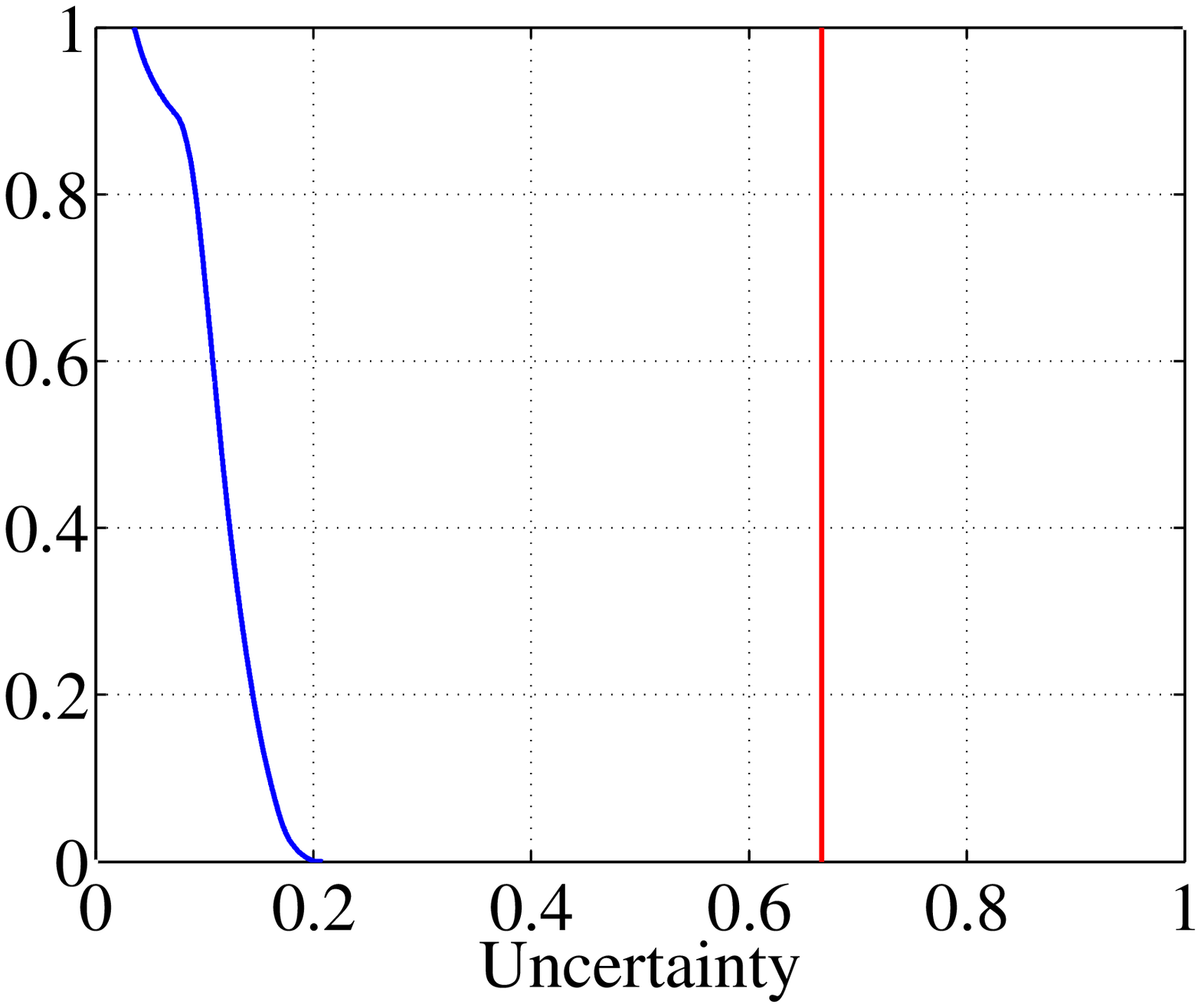}~  
 \includegraphics[clip =true, trim = 0in 0.39in 0in 0in, width=0.45\columnwidth]{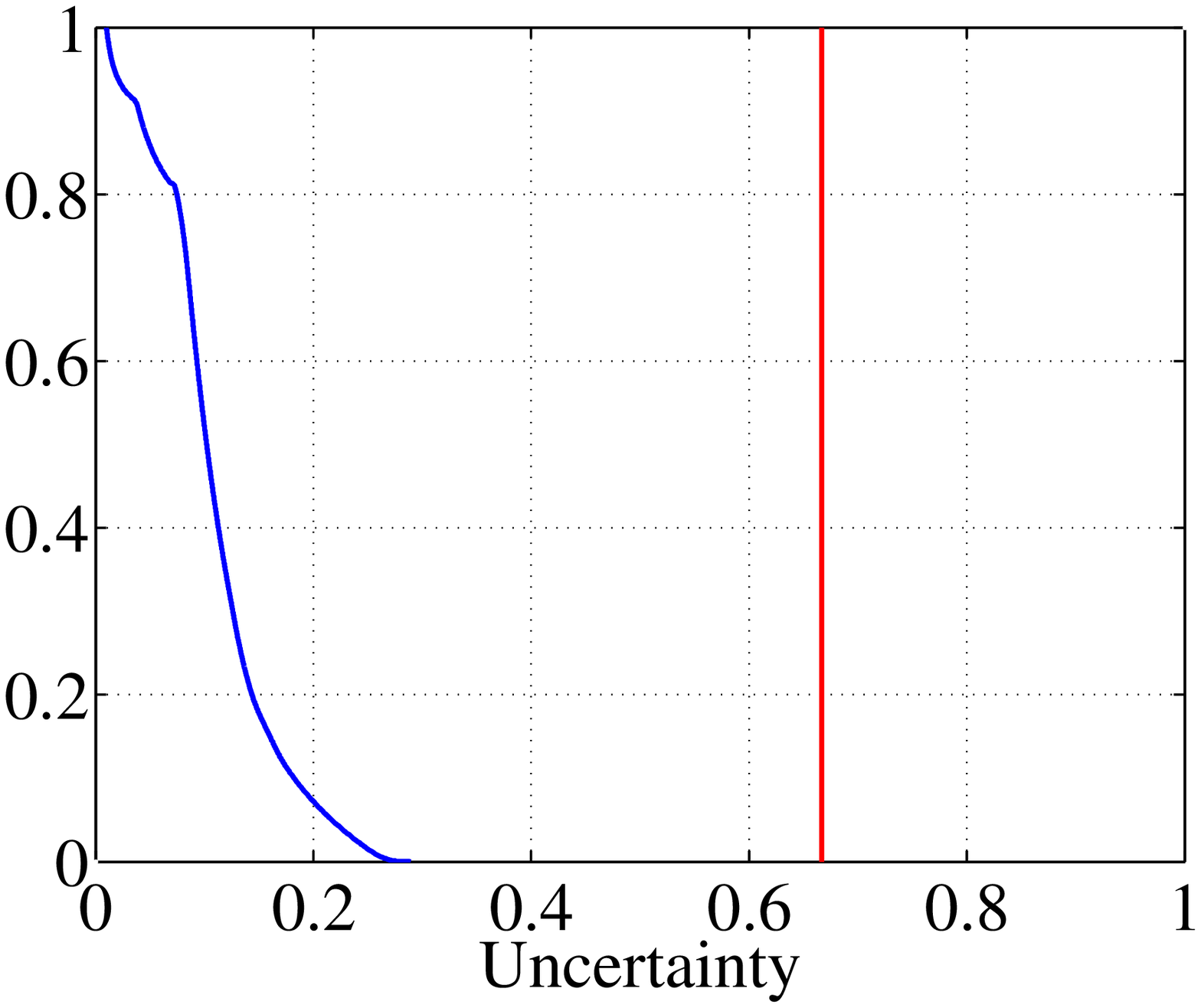}

 \caption{CCDF of uncertainty seen by packets as they cut through the network (after bootstrap) for $\epsilon=0$ (left) and $\epsilon = 2$ (right). The red vertical line demarcates the upper bound on uncertainty \eqref{eq:route_quality_bound}}

 \label{fig:uncertainty_pkt}

 \end{figure}

\separation
\noindent {\bf Uncertainty:} The uncertainty of the position estimate is designed to be less than $\left.\alpha\beta\middle/\left(1-\beta\right)\right.=2/3$ once the packet bootstraps. From our simulations, we find that the uncertainty seen by a packet after bootstrap remains smaller than this value. The CCDFs of the location uncertainty after bootstrap are presented in Figure~\ref{fig:uncertainty_pkt}.

\begin{figure}

\centering

\includegraphics[clip =true, trim = 0in 0.39in 0in 0in, width=0.45\columnwidth]{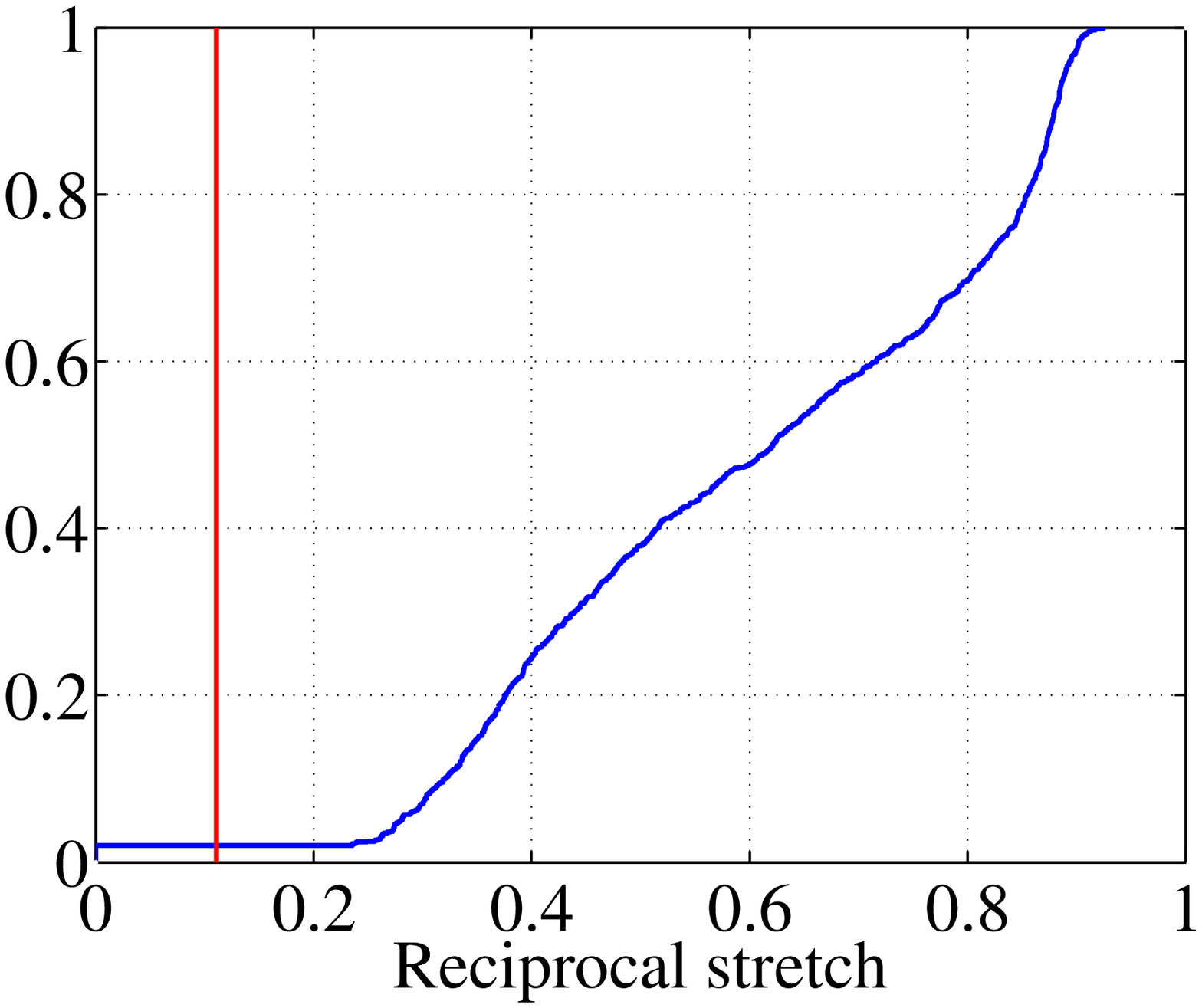}~
\includegraphics[clip =true, trim = 0in 0.39in 0in 0in, width=0.45\columnwidth]{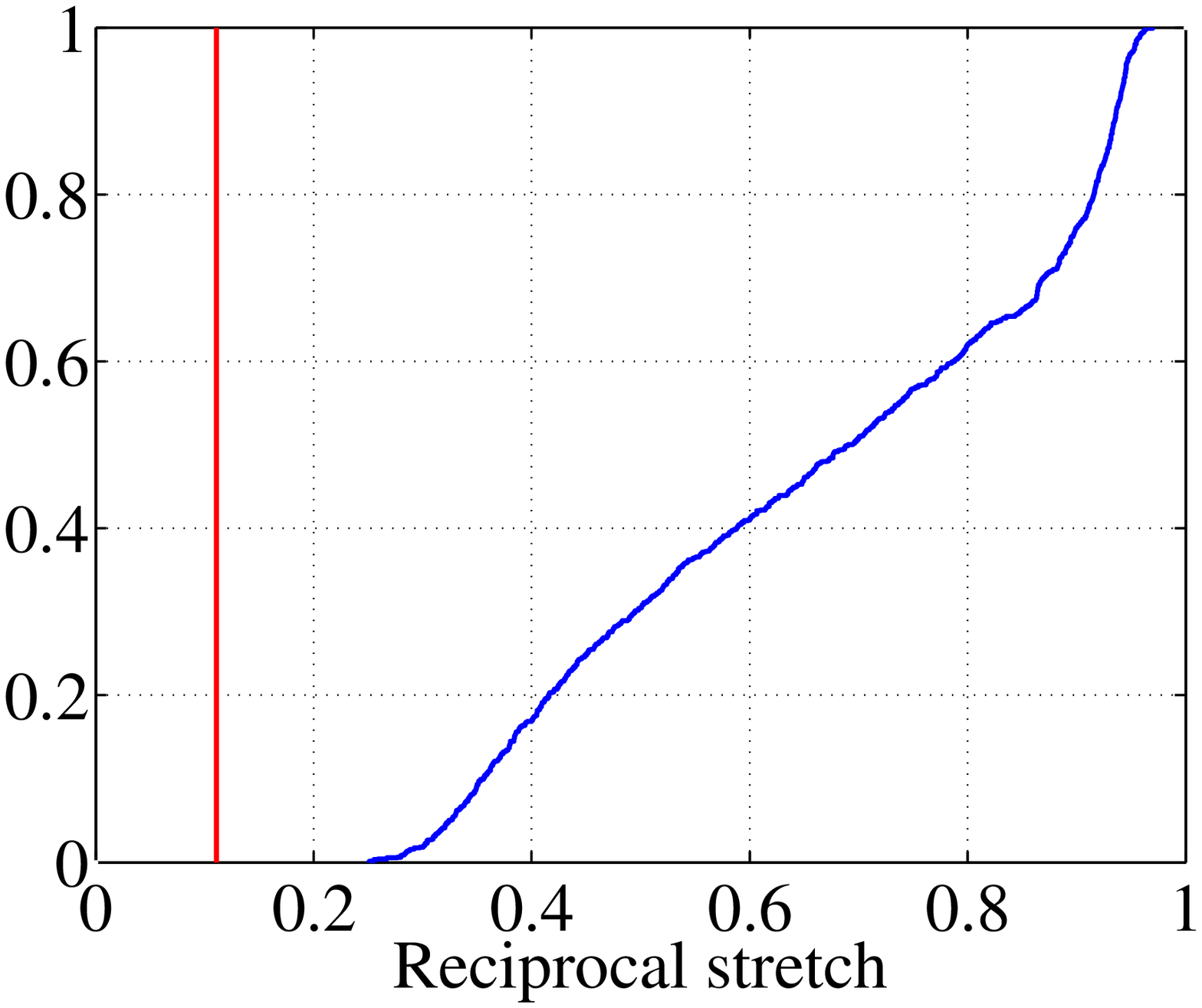}

\caption{CDF of reciprocal stretch, i.e., the ratio of source-destination distance to the length of the packet trajectory, for communication radius $r$ corresponding to $\epsilon=0$ (left) and $\epsilon=2$ (right). Red vertical line corresponds to the worst case stretch (\ref{eq:stretch_expression})}

\label{fig:reciprocal_stretch}

\end{figure}

\separation
\noindent {\bf Route Stretch:} Figure~\ref{fig:reciprocal_stretch} plots the CDF of the {\it reciprocal} of the route stretch attained, with reciprocal stretch equal to zero indicating a routing failure (edge effects are ignored by focusing on trajectories which start inside the outermost update ring). From \eqref{eq:stretch_expression}, the route stretch is bounded by $9$ and the corresponding reciprocal stretch is marked via the red vertical line. Note that all successful routes satisfy this guarantee, and that all routes are successful for $\epsilon = 2$. For $\epsilon = 0$ (greedy geographic forwarding not guaranteed to work), a small fraction of route failures do occur.

\section{Conclusions}  \label{sec:conclusions}

To the best of our knowledge, this is the first work that provides a provably scalable position-publish protocol while providing guarantees on route stretch for the accompanying geographic  routing protocol. Key to scalability is a probabilistic approach to updating a {\it subset} of nodes, and to geographic forwarding with imperfect information. Our emphasis here was on providing analytical insight and design criteria, verified by simulations. Mapping our ideas to practice require detailed protocol specifications at the level of packet level format and processing, and more extensive simulations for a wide variety of mobility models. In addition, while we focus on distant nodes in proving scalability, it may be possible to significantly optimize our protocol as the distance to the destination decreases. Another direction for future investigation is the design of position-publish strategies that account for large holes in the deployment region. Finally, it is interesting to note that, while we have assumed a uniform set of protocol parameters for all destination nodes to prove scalability, in practice, each potential destination can choose its parameters differently, depending on the tradeoffs between routing overhead, reliability and stretch that it desires to obtain.

\appendices

\section{Proof of Theorem~\ref{thm:relaying:accuracy}}{ \label{appendix:proof:lemma:relay:accuracy}

We start by laying out the notation used in this proof. Every node in the network can forward packets to any node that is within the communication radius $r(n)$. We refer to the circle of radius $z$ centered around a point $\bu$ by $C(\bu,z)$. We reserve the symbol $r=r(n)$ for the communication radius and $\bv$ for the location of the network node (relay) under consideration. Therefore, the node $\bv$ can forward a packet to any node inside the circle $C(\bv,r)$.

\begin{figure}

\centering

\includegraphics[clip =true, trim=0.825in 0in 0in 0.825in, width=0.85\columnwidth]{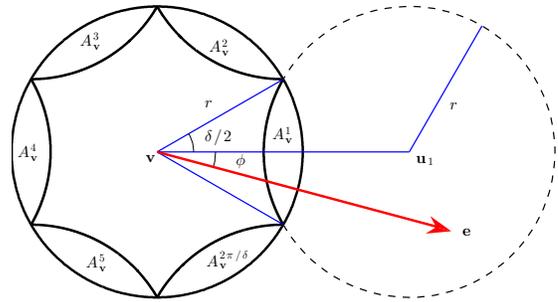}

\caption{Anchor regions $A_{\bv}^{k}$ numbering $\lceil 2\pi/\delta\rceil$ in the neighborhood (distances smaller than the communication radius $r$) around a typical node $\bv$. $\phi$ is the direction of the estimate $\be$ towards which a packet is being greedily routed by $\bv$.}

\label{fig:union:bound:anchors}

\end{figure}

Around each node $\bv$, we choose $\lceil 2\pi/\delta\rceil$ ``anchor'' regions $A_{\bv}^{k}$ of the following form: (i) $A_{\bv}^{k} = C(\bv,r)\cap C\left(\bu_k,r\right)$, where $\bu_k$ is a point on the perimeter of the circle $C\left(\bv,2r\cos\left(\delta\middle/2\right)\right)$, where $\delta$ is the constant in the statement of Theorem~\ref{thm:relaying:accuracy} (satisfying $0 < \delta \leq \pi/3$) (ii) The $\lceil 2\pi/\delta\rceil$ points $\{\bu_k\}$ are carefully chosen so that any ray drawn outward from this node $\bv$ intersects at least one anchor region $A_{\bv}^{k}$. Figure~\ref{fig:union:bound:anchors} illustrates one such choice of anchor regions around the node $\bv$ for $\delta = \pi/3$ ($2\pi/\delta$ chosen to be an integer for the sake of convenience).

We now provide a summary of the proof strategy used. Using an union bound, we first show that when we choose a large enough communication radius ($ r^2 \geq c \log n$, for an appropriate choice of constant $c$), then w.h.p., every anchor region in the network is occupied by at least one node. We follow this up with the implications of the occupancy of the $\left\lceil2\pi\middle/\delta\right\rceil$ anchor regions $\{A_{\bv}^{k}\}$ around a node $\bv$ for greedy forwarding decisions made by this node.

\subsection{Asymptotic occupancy guarantee for anchor regions }

Let $E_{\bv}^{k}$ denote the ``error event'' that all $n-1$ network nodes (other than $\bv$) reside outside the anchor region $A_{\bv}^{k}$. In other words, $E_{\bv}^{k}$ denotes the event that the anchor region $A_{\bv}^{k}$ is unoccupied. All nodes are uniformly and independently distributed over the deployment region of area $n$. Therefore, denoting the area of the region $P$ by $|P|$, we have that
\begin{equation*}
\Pr\left[E_{\bv}^{k}\right]  =  \left(1 - \left.{\left|A_{\bv}^{k}\right|}\middle/{n}\right.\right)^{n-1}\!\!.
\end{equation*}
We refer to the event that one of the $n\left\lceil {2\pi}\middle/{\delta} \right\rceil$ anchor regions being empty as the ``cumulative error event'' and denote it by $\Eone$. Since $\Eone = \bigcup_{\bv,k} E_{\bv}^{k}$, we use the union bound to arrive at:
\begin{equation*}
\Pr\left[\Eone\right]  \leq \sum_{\bv, k} \Pr\left[E_{\bv}^{k}\right] = \left\lceil \frac{2\pi}{\delta} \right\rceil n \left(1 - \frac{\left(\delta - \sin\delta\right)r^2}{n}\right)^{n-1}\!\!\!\!\!\!\!\!\!,
\end{equation*}
where we have used the fact that $\left|A_{\bv}^{k}\right| = \left(\delta - \sin\delta\right)r^2$. Let $\nu>0$ be a some constant. We note that when
\begin{equation}
r^2 = \left.\left(1+\nu\right)\log n\middle/\left(\delta - \sin\delta\right)\right.,
\label{eq:minimum_radius_proof_lemma}
\end{equation}
$\Pr\left[\Eone\right] \rightarrow 0$ as $n$ grows. Thus, for large $n$, if $r$ satisfies \eqref{eq:minimum_radius_proof_lemma}, all anchor regions are occupied w.h.p.

\subsection{Implications of anchor region occupancy for greedy forwarding}

Consider a packet at the node $\bv$, which is being forwarded towards an estimate $\be$ (need not correspond to the location of any of the $n$ nodes in the network) which is no closer to $\bv$ than $2r$. All $\lceil 2\pi/\delta\rceil$ anchor regions $\{A_{\bv}^k\}$ around $\bv$ are occupied by at least one node. We denote the line segment joining the points $\bu$ and $\bv$ by $L(\bu,\bv)$. w.l.o.g., we assume that $L(\bv,\be)$ intersects the anchor region $A_{\bv}^{1}$. Let $\phi$ denote the angle between $L(\bv,\be)$ and $L(\bv,\bu_1)$, where $\bu_1$ is the point used in the construction of the anchor region $A_{\bv}^{1}$ (this is depicted in Figure~\ref{fig:union:bound:anchors}). We now provide the answer to the question: What does the occupancy of the region $A_{\bv}^1$ mean for the quality of the greedy forwarding decision taken by $\bv$ corresponding to this estimate $\be$? More specifically, denoting the next hop (the node to which this packet is forwarded to) by $\bx$, we ask, what is largest absolute value that the angle between $L(\bv,\be)$ and $L(\bv,\bx)$ can take?

\begin{figure}

\centering

\includegraphics[clip=true, trim=1.6in 0.2in 0in 1in, width=0.65\columnwidth]{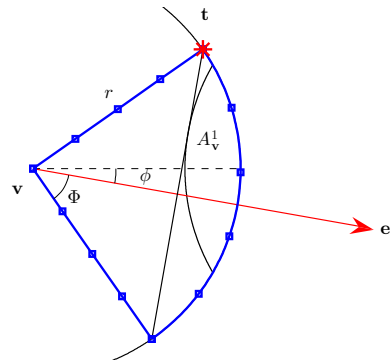}

\caption{Sector $S$ is marked using the {\tiny $\Box$} symbol. The point $\bt$ in $C(\bv,r)\setminus S$ closest to the estimate $\be$ is marked using the $*$ symbol.}

\label{fig:geometric:implication:anchors}

\end{figure}

Consider the tangent to the inner boundary of the anchor region $A_{\bv}^1$ which is perpendicular to the line segment $L(\bv,\be)$ (tangent to $C(\bu_1,r)$; marked in Figure~\ref{fig:geometric:implication:anchors}). Let $2\Phi$ denote the angle subtended by the acute sector $S$ of $C(\bv,r)$ associated with this tangent  (we mark this sector in Figure~\ref{fig:geometric:implication:anchors} using the {\tiny $\Box$} symbol). It can be shown that 
$
\cos\Phi = 2\cos\left(\delta/2\right)\cos\phi - 1.
$
Since the line $L(\bv,\be)$ intersects this anchor region $A_{\bv}^1$, we have that $\phi\leq\delta/2$, and as a result $\Phi\leq\delta$.

Consider any point $\bu \in C(\bv,r)\setminus S$.  When $\delta\leq \pi/3$ and $\ell(\bv,\be)> 2r$, it can be shown that: (i) $\ell(\bu,\be)\geq\ell(\bt,\be)$ (where $\bt \in C(\bv,r)\setminus S$ is the point marked in Figure~\ref{fig:geometric:implication:anchors} using the $*$ symbol) (ii) $C(\be,\ell(\be,\bt))$ encompasses the anchor region $A_{\bv}^{1}$ completely. Therefore, every point inside the anchor region $A_{\bv}^1$ is closer to $\be$ than any point in $C(\bv,r)\setminus S$ (i.e., $A_{\bv}^1\subset C(\be,\ell(\be,\bt)) \subseteq C(\be,\ell(\be,\bu))$).

We now show (by contradiction) that when $\delta < \pi/3$, $\ell(\bv,\be) > 2r$, the next hop $\bx$ lies within this sector $S$ of the of width $2\Phi$. Now suppose that the next hop $\bx\in C(\bv,r) \setminus S$. Our preceding discussions imply that $A_{\bv}^1\subset  C(\be,\ell(\be,\bx))$. Since greedy geographic forwarding always chooses the neighbor closest to the destination, we infer that the anchor region $A_{\bv}^1$ is not occupied. But this contradicts the assumption that $A_{\bv}^1$ is occupied by at least one node. Thus, the next hop $\bx$ must lie inside the sector $S$. The absolute angle between direction along which the packet is forwarded and the direction of the estimate using which it is forwarded is therefore bounded by $\Phi\leq\delta$.

\subsection{Summary}
We have shown that: (i) when the communication radius $r(n)$ scales as \eqref{eq:minimum_radius_proof_lemma}, then w.h.p., all anchor regions around every node are occupied. This corresponds to choosing $\epsilon$ in \eqref{eq:comm_rad_choice} so that \eqref{eq:slack:comm:rad:imperfect:estimates} holds (ii) When all the $\left\lceil2\pi\middle/\delta\right\rceil$ anchor regions around the node $\bv$ are occupied, the absolute angle between the direction along which the packet is forwarded $L\left(\bv,\bx\right)$ and the direction of the position estimate $L\left(\bv,\be\right)$ using which this forwarding decision is made can be no larger than $\delta$ (for any $\delta < \pi/3$ and $\be$ such that $\ell\left(\bv,\be\right) > 2r$), thus proving Theorem~\ref{thm:relaying:accuracy}.  \qed



\subsection{Proof of Corollary~\ref{corr:uncertainty:reliability}}

Consider the scenario where the uncertainty of all position estimates is bounded by $\Umax$. This ensures that the angle between the estimated direction and the true direction is no greater than $\arcsin \Umax$. If the communication radius $r$ is large enough, we infer from Theorem~\ref{thm:relaying:accuracy} that greedy forwarding decisions result in angular distortions smaller than $\delta$ (as long as the relay-estimate distance is no smaller than $2r$). Therefore, the angle between the direction along which the packet is forwarded and the true direction is bounded by $\arcsin\left(\Umax\right) + \delta$ and progress towards the destination per unit distance travelled is no lesser than $\cos\left(\arcsin\left(\Umax\right) + \delta\right)$. Therefore, every hop reduces the packet-destination distance until the relay-estimate distance is smaller than $2r$. From the definition of uncertainty, we have that when the relay-estimate distance is smaller than $2r$, the relay-destination distance is smaller than $2\left(1-\Umax\right)^{-1} r$. Therefore, location information is exact and routing is guaranteed to be successful thereafter (when $\epsilon$ satisfies \eqref{eq:slack:comm:rad:imperfect:estimates} for some $\delta\leq\pi/3$, $\epsilon>\epsilon_0\approx 1.6$ needed for successful routing with exact location information \cite{AsymptoticCriticalGreedyGeo:TC}).

}


\section{Bound on probability of missing an update ring}{\label{sec:appendix:Pmiss}

The density of nodes throughout the network is a uniform $1$ node per unit area. This remains invariant under our Brownian motion model. For $r_i \leq a \leq r_i + d_i$, let us denote by $\Lambda_U ( a,t)$ the ``update density'', or the density of the subset of nodes in the ring with active updates, where $t$ is the time elapsed since update issue and $a$ is the distance of from the center of the update ring (the update density is circularly symmetric and thus is a function of only the distance from the center of the ring). At $t=0$, all nodes in the ring have active updates, so that $\Lambda_U (a,0) = I_{[r_i,r_i+d_i]} (a)$, where $I_B$ denotes the indicator function of a set $B$.  As time proceeds, the positions of the nodes with active updates is smeared out by the Gaussian kernel induced by 2D Brownian motion, so that 
\begin{equation}
\Lambda_{U}\left(a,t\right)=\Lambda_{U}\left(a,0\right)\otimes\mathcal{N}\left(\mathbf{0},\sigma^{2}t~\mathbb{I}_{2}\right) 
 \label{eq:2dconv} 
\end{equation} 
where $\otimes$ stands for $2$D convolution. Let $\Lambda^{\star} (a) = \Lambda_U (a,T_i)$ be the worst case update density (just before the update expires).  When a packet meets a node at a distance $a$ from the center of the $i$-th update ring, the probability that it does not get an active update is therefore at most $1 - \Lambda^{\star} (a)$.

The worst case packet traversal for missing an update ring is given by a radial cut through, and for this trajectory the packet meets at least $\left.{d_i}\middle/{r}\right.$ nodes, and a miss occurs if none of these have an active update (we wish to reiterate that meeting fewer nodes inside an update region increases the chance that the packet misses this update and that the number of nodes via which a packet is relayed can be no lesser than $d_i/r$). We therefore obtain that the miss probability for the $i$-th ring satisfies 
\begin{equation}
\Pmiss (i) \leq \prod_{\ell =1}^{\ell = \left.{d_{i}}\middle/{r}\right.} \left(1-{\Lambda^{\star} \left( r_i + \ell r\right)}\right). 
\label{eq:pmiss_min_hops} 
\end{equation}
Taking logarithms, the product becomes a sum which we then approximate as an integral using $r/d_i \rightarrow 0$ for $i$ large. 
\begin{align} 
\log \Pmiss(i) & \leq  \sum_{l=1}^{l =\left.{d_{i}}\middle/{r}\right.}\log\left(1-{\Lambda^{\star} \left( r_i + \ell r\right)}\right)\nonumber \\ 
&\leq - \sum_{l=1}^{l =\left.{d_{i}}\middle/{r}\right.}{\Lambda^{\star}\left(r_{i}+lr\right)}\nonumber \\ 
& \approx  -\frac{1}{ r}\int_{r_{i}}^{r_{i}+d_{i}}{\Lambda^{\star}\left(a\right)}~\da .
\label{eq:log_pmiss} 
\end{align}

The worst case update density for the $i$-th ring $\Lambda^{\star}(a)$ ($a$ is the distance from the center of the update ring) is the density just before the timer $T_i$ corresponding to the update elapses and from (\ref{eq:2dconv}), we have that
\begin{equation}
 \Lambda^{\star} \left(a\right) = \frac{1}{\sigma^{2}T_{i}}\overset{r_{i}+d_{i}}{\underset{r_{i}}{\int}}\rho\exp\left(\frac{-a^{2}-\rho^{2}}{2\sigma^{2}T_{i}}\right)I_{0}\left(\frac{a\rho}{\sigma^{2}T_{i}}\right) \drho,
 \label{eq:worst_case_density}
 \end{equation}
where $I_{0}\left( \cdot \right)$ denotes the zeroth order modified Bessel function of the first kind. The probability of missing the the $i$-th update ring satisfies \eqref{eq:log_pmiss}. Using \eqref{eq:worst_case_density} in \eqref{eq:log_pmiss},
\begin{align*}
\log \Pmiss(i) & \lessapprox \frac{-1}{\sigma^{2}rT_{i}} {\iint_{\mathcal{A}_i}} \rho \exp\left({\frac{-a^2-\rho^2}{2\sigma^{2}T_{i}}}\right)I_0\left(\frac{a\rho}{\sigma^2T_i}\right)\da\drho,
\end{align*}
where $\mathcal{A}_i$ is given by $r_{i}\leq a,\rho \leq \left(r_i+d_i\right)$. For outer rings $\min_{\mathcal{A}_i} \left(a\rho\right) = r_i^2 \gg \sigma^2 T_i$ (since $\gamma < 2$; required to limit abnormal updates). Therefore, the argument of $I_0(\cdot)$ is large and we can use the approximation $I_0(t) \geq \left.\exp(t)\middle/\sqrt{2\pi t}\right.$ (which is valid for large $t$)  to arrive at
\begin{align*}
\log \Pmiss(i) & \lessapprox \frac{-\sigma\sqrt{T_i}}{r\sqrt{2\pi}} {\iint_{\mathcal{B}_i}} \sqrt{y/x} \exp\left(-\frac{\left(x-y\right)^{2}}{2}\right)\dx~\dy,
\end{align*}
where we have set $x = \left.a\middle/\sqrt{\sigma^2 T_i}\right.$, $y = \left.\rho\middle/\sqrt{\sigma^2 T_i}\right.$ and $\mathcal{B}_i$ is given by $\left.r_{i}\middle/\sqrt{\sigma^2 T_i}\right.\leq x,y \leq \left.\left(r_i+d_i\right)\middle/\sqrt{\sigma^2 T_i}\right.$. We note that $\min_{\mathcal{B}_i}\sqrt{y/x} = \sqrt{\left.r_i\middle/\left(r_i+d_i\right)\right.}$. Since $r_i$ scales faster than $d_i$, this is well-approximated by $1$ for outer rings. Using this approximation, we arrive at
\begin{align}
\log \Pmiss(i) & \lessapprox \frac{-\sigma\sqrt{T_i}}{r\sqrt{2\pi}} {\iint_{\mathcal{B}_i}} \exp\left(-\left(x-y\right)^{2}\middle/2\right)\dx~\dy\nonumber\\
& = \frac{-\sigma\sqrt{T_i}}{r} \times \int_0^{\deff_i} \left( 1 - 2 Q\left(x\right)\right)~\dx \nonumber\\
 &\approx \frac{-\sigma\sqrt{T_i}}{r\sqrt{2\pi}} \times \left(\deff_i\right)^2.\nonumber
\end{align}
where $Q(x)$ is the CCDF of the standard normal distribution given by $\left(1\middle/\sqrt{2\pi}\right)\int_x^{\infty} e^{-t^2/2}~\textrm{d}t$ and $\deff_i$ denotes $\left.{d_i}\middle/\left(\sigma\sqrt{T_i}\right)\right.$. The above approximation is accurate when $\deff_i$ is small enough so that for $0\leq x\leq \deff_i$, $Q(x)$ is well approximated by $0.5-\left(x\middle/\sqrt{2\pi}\right)$. As we choose $\gamma > 1+\mu$ to accommodate scalability constraints, this is equivalent to the ring index $i$ being large enough. Thus, for outer rings, 
$
\log \Pmiss(i) \lessapprox \left. -d_i^2\middle/\left(r\sigma\sqrt{2\pi T_i}\right)\right..
$

}

\section{Bound on route stretch}{ \label{sec:appendix:route:stretch}

To provide bounds on the worst case route stretch, we need to understand the geometry of the update rings surrounding a destination node that executes the position-publish algorithm. The publish algorithm ensures that at all times, there exists \emph{exactly one set} of normal updates with valid confidence region guarantees corresponding to each ring index. All other updates with valid confidence region guarantees are abnormal updates issued in order to prevent older unexpired updates (made stale by atypically large movements of the destination node) from misdirecting packets. When a packet is relayed through an update ring to which updates that were issued earlier have become stale, abnormal updates issued to the same region ensure that the packet latches on to the newer estimate that they possess rather than the stale estimates. This is because the routing algorithm prefers newer updates (more recent) when it is presented with a tie in terms of the ring indices. The analysis in the derivation of $\Pmiss\left(i\right)$ holds here for the probability of missing the \emph{newer} updates (in regions with stale updates that also have the same spatial validity). The role of abnormal updates is therefore to prevent routing failures as a result of misdirection from stale updates. Therefore, we can assume that abnormal updates and stale updates which were compensated for using abnormal updates are absent in a discussion of worst case routing stretch guarantees.

Consider a destination node which is surrounded by one normal update ring of each ring index, all of which satisfy their corresponding confidence region guarantees. We denote the center and the position estimate (both of which coincide for normal updates) of the $i$-th update ring by $\bc_{i}$. Since these position updates satisfy their corresponding confidence region guarantees, $\ell\left(\bc_{i},\bd\right)\leq\beta r_{i}$.

\separation
\noindent\textbf{Inner and outer envelope of updates:}  In the following computations it is useful to define the concept of inner and outer envelope of an update of ring index $i$. The outer envelope of updates of ring index $i$ is defined as the set of points farthest from the destination node that can possess a spatially active update of ring index $i$. Similarly the inner envelope of updates of ring index $i$ is defined as the set of points closest to the destination node that can possess a spatially active update of ring index $i$. In each direction the farthest spatially valid update can be $\left(1+\beta\right)r_{i}$ away corresponding to $\ell\left(\bc_{i},\bd\right) =\beta r_{i}$ and $\bc_{i}$ in the same direction (we neglect $d_{i}$ in this computation as $\left.{d_{i}}\middle/{r_{i}}\right.\ll1$). So the outer envelope of updates of ring index $i$ is the circle of radius $\left(1+\beta\right)r_{i}$ centered around the destination node $\bd$. Similarly the inner envelope of updates is the circle of radius $\left(1-\beta\right)r_i$ centered around the destination node.

If the packet source has an active estimate of the destination, then the packet proceeds towards this estimate. However, if it does not possess an active update, the source launches the packet in some arbitrarily chosen direction and the packet eventually \emph{bootstraps} when it is relayed to a node which possesses an active estimate of the destination node's location. This process of bootstrapping contributes to route stretch in addition to the route stretch stemming from lazy updates.

Firstly, we examine the contribution of the lazy position updates to the route stretch after bootstrap.  An upper bound on this is given via the worst case uncertainty seen by a packet that has bootstrapped. Then we provide an upper bound on the worst case contribution to route stretch due to the bootstrapping process which corresponds to a confluence of unfavorable geometric configurations of the immediate inner and outer rings surrounding the packet source and the launch direction chosen by it.

\subsection{Worst cast stretch after bootstrap}

What is the worst case uncertainty seen by a packet after it has bootstrapped?  To answer this question consider a segment of any packet trajectory from just after it has acquired the update for the $(i+1)$-th ring till it acquires the update for the $i$-th ring. In this segment of the packet trajectory, the packet is routed using the estimate from the $(i+1)$-th ring and so the location estimate $\be = \bc_{i+1}$ can disagree from the true destination location by not more than $\beta r_{i+1}=\alpha\beta r_{i}$ (i.e., $\ell\left(\bd,\be\right)\leq \beta r_{i+1} = \alpha\beta r_i$). The farthest from the destination node that an update of index $i+1$ can be obtained by the packet is given by $\left(1+\beta\right)r_{i+1}$, the radius of the outer envelope of the $(i+1)$-th ring and the closest the packet can get to the destination without latching on to the $i$-th ring update is $\left(1-\beta\right)r_{i}$, the radius of inner envelope of updates of the $i$-th ring. This is the region where a packet can use the $(i+1)$-th indexed update, and in this region, uncertainty ${\ell\left(\bd,\be\right)}/{\ell\left(\bp,\bd\right)}$ satisfies $U\leq \Umax$, where
\begin{equation} 
\Umax = \left.\alpha\beta\middle/\left(1-\beta\right)\right..
\label{eq:route_quality_bound}
\end{equation}
The above bound is independent of the update index $i+1$ and thus as long as the packet does not ``miss'' any update ring that it is relayed through, the uncertainty seen by it is no greater than $\Umax$. Therefore, from Theorem~\ref{thm:relaying:accuracy}, the route stretch after bootstrap is at most $1/\cos\left(\arcsin\Umax + \delta\right) \approx 1/\sqrt{1-\Umax^2}$ (assuming that $\delta$ is small and $\epsilon$ in \eqref{eq:comm_rad_choice} is chosen to satisfy \eqref{eq:slack:comm:rad:imperfect:estimates}). Therefore, when $\Umax = \left.\alpha\beta\middle/(1-\beta)\right. < 1$, the route stretch after bootstrap is bounded.

From the preceding discussion, we see that when $\Umax < 1$ or $\beta r_{i+1} < (1 -\beta)r_i $, before the packet reaches the estimate $\bc_{i+1}$ corresponding to the $(i+1)$-th ring, it is guaranteed to be relayed via the $i$-th ring, thus acquiring the estimate $\bc_i$ corresponding to the $i$-th update. Therefore, the packet is guaranteed to progressively obtain better estimates of the destination, and this imposes the extra constraint on routing reliability (Section~\ref{sec:reliability}) given by $\alpha\beta < 1- \beta$ or $\Umax < 1$.

\subsection{Bootstrapping cost}

Consider a packet originating from the source node positioned at $\bs$ for the destination at $\bd$. Suppose the rings that surround the source are the $i$-th (inner) and $(i+1)$-th (outer) rings. A packet launched in any direction first bootstraps at one of these two rings. Since we are interested in route stretch, which is a ratio of distances, we henceforth scale all distances by $r_{i}$, the radius of the inner ring. The source can be surrounded by the $i$-th and $(i+1)$-th ring only if it is inside the region between the inner envelope of the $i$-th ring and the outer envelope of the $(i+1)$-th ring. Therefore $1-\beta\leq {\ell\left(\bs,\bd\right)}\leq\alpha\left(1+\beta\right)$.

When the packet is launched from the source in an arbitrary direction, the scenario where the packet latches to the ring closer to the destination is a better case. So consider the case where the packet bootstraps at the outer ring. We note that for all distances of the source node from the destination node the packet travels the farthest distance before bootstrapping if it is launched tangential to the inner envelope of the inner ring and bootstraps at the outer envelope of the outer ring. This scenario is depicted in Figure~\ref{fig:Stretch-computations} (left). We denote the point where this tangential trajectory touches the inner envelope of the inner ring by $\bt$ and the point on the outer envelope of the outer ring where the packet bootstraps by $\mathbf{b}$.

\begin{figure}

\centering

\includegraphics[clip=true,width=0.425\columnwidth]{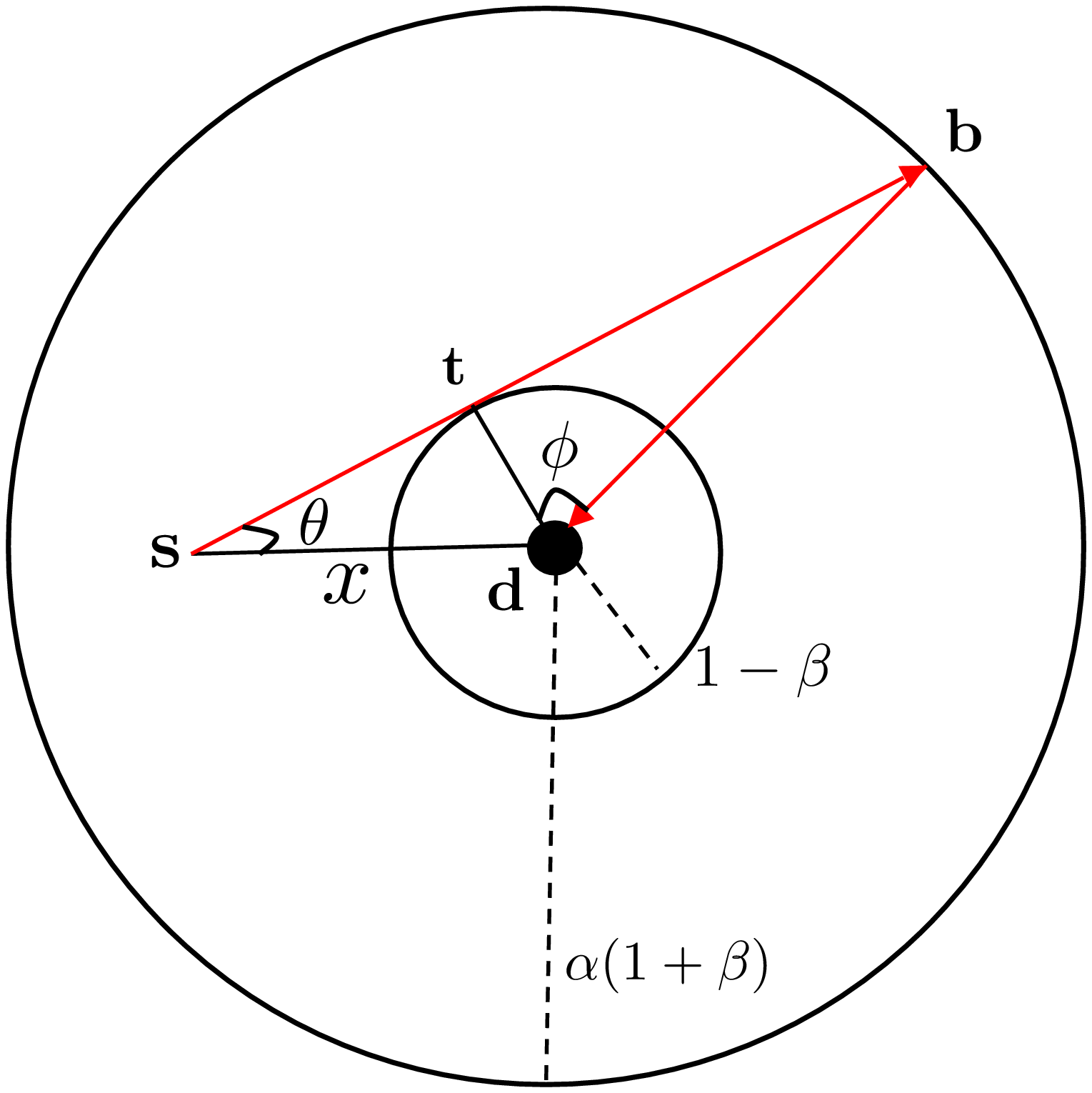} \quad 
\includegraphics[clip=true,width=0.38\columnwidth]{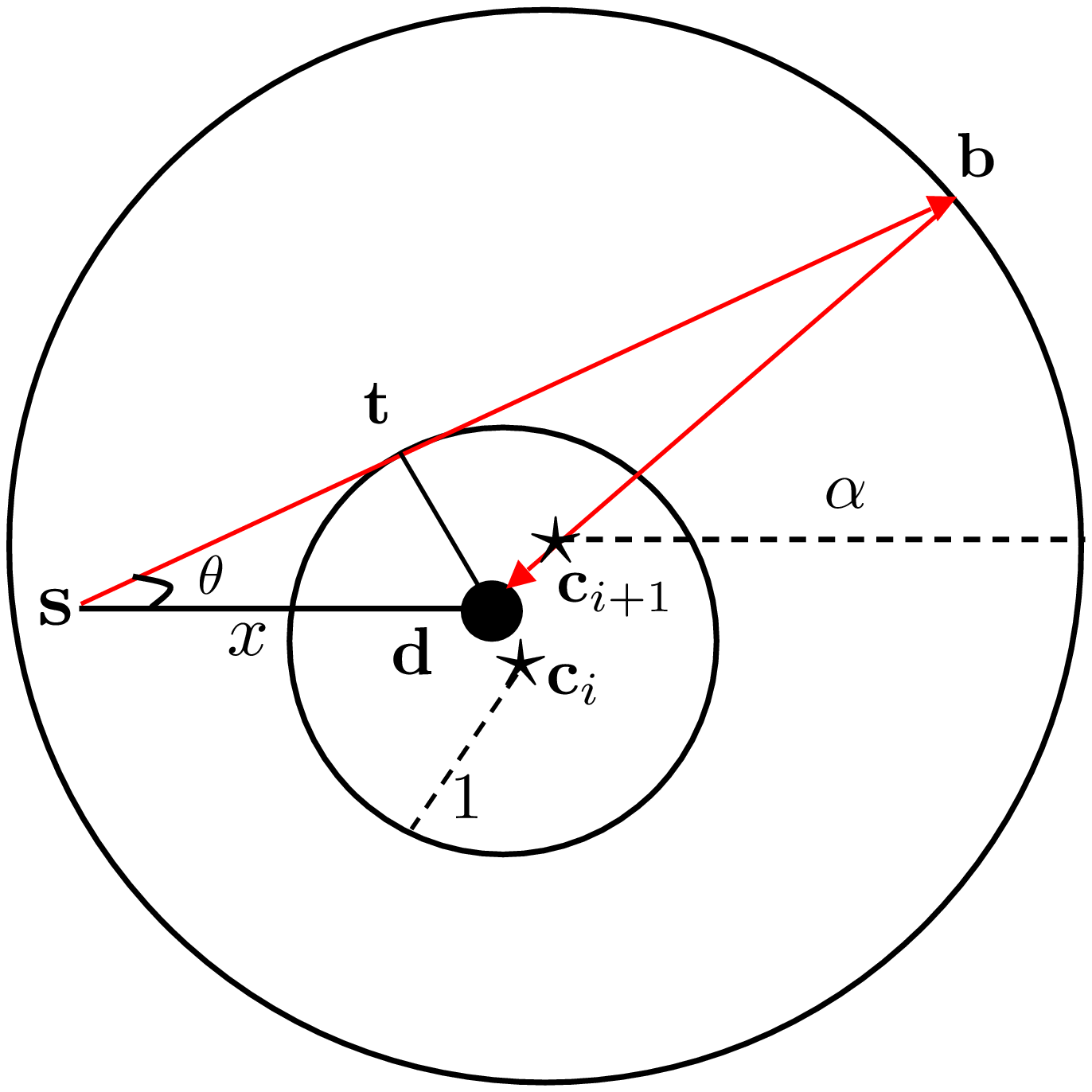}

\caption{Left: Contribution of bootstrapping to stretch via envelopes. Right: A configuration of ring centers $\bc_{i}$ and $\bc_{i+1}$ marked $\star$ which satisfies the confidence region guarantees and has the same bootstrapping cost as the worst case envelope based calculations}

\label{fig:Stretch-computations}

\end{figure}

Let $\theta$ be the angle between the launching direction and the direction of the destination. Then $\sin\theta={(1-\beta)}/{x}$.  Let $\phi$ be the angle between the vectors $\bt$ and $\mathbf{b}$ as shown in Figure~\ref{fig:Stretch-computations} (left). Then we have $\cos\phi={(1-\beta)}/{\left(\alpha\left(1+\beta\right)\right)}$. From the cosine formula, the worst case distance traveled before bootstrap $Z(x)=\ell\left(\bs, \mathbf{b}\right) $ for a packet originating at a distance $x = \ell\left(\bs,\bd\right)$ from the destination satisfies (using Theorem~\ref{thm:relaying:accuracy} and assuming that $\epsilon$ is large enough so that $\delta$ in \eqref{eq:slack:comm:rad:imperfect:estimates} is small; note that forwarding along a fixed direction is equivalent to forwarding greedily towards an estimate at $\infty$ and thus in the regime where Theorem~\ref{thm:relaying:accuracy} is applicable)
\begin{align*} 
Z(x) &= \sqrt{x^{2}+\alpha^{2}\left(1+\beta\right)^{2}-2x\alpha\left(1+\beta\right)\cos\left(\frac{\pi}{2}-\theta+\phi\right)}\\ 
& = \sqrt{x^{2}-\left(1-\beta\right)^{2}}+\sqrt{\alpha^{2}\left(1+\beta\right)^{2}-\left(1-\beta\right)^{2}}.
 \end{align*}

 \subsection{Bound on overall stretch}

We are now ready to bound the overall stretch using the preceding two ingredients: A packet originating at a distance $x$ from the destination travels a distance not exceeding $Z(x)$ before bootstrap at $\mathbf{b}$. Since the stretch after bootstrap is at most $1/{\sqrt{1-\Umax^2}}$ ($\Umax$ is the upper bound on uncertainty given by \eqref{eq:route_quality_bound}) and $\ell\left(\mathbf{b},\bd\right) = \alpha(1+\beta)$, the overall route stretch is at most $\texttt{S}$:
\begin{align*} 
\texttt{S} & =  \max_{\ell\left(\bs,\bd\right)}  \left.\left(\ell\left( \bs,\mathbf{b}\right) + \left(\ell\left(\mathbf{b},\bd\right)\middle/\sqrt{1-\Umax^2}\right)\right)\middle/{\ell\left(\bs,\bd\right)}\right.\\
& = \max_{1-\beta\leq x\leq\alpha\left(1+\beta\right)}\Bigg({\sqrt{x^{2}-\left(1-\beta\right)^{2}}}+\frac{\alpha\left(1-\beta^2\right)}{\sqrt{(1-\beta)^2-{\alpha}^2{\beta}^{2}}}\\ 
&  \quad\quad\quad\quad\quad\quad\quad\quad +{\sqrt{\alpha^{2}\left(1+\beta\right)^{2}-\left(1-\beta\right)^{2}}}\left.\Bigg)\middle/x\right.
\end{align*}
and it can be shown that this maximum is equal to the expression in (\ref{eq:stretch_expression}).

 While we consider worst case envelopes of rings for the above discussion, this scenario can be mapped to a feasible configuration of inner and outer update rings as is shown in Figure~\ref{fig:Stretch-computations} (right) because each point on the envelopes corresponds to a certain valid choice of ring center. The center of the inner ring is $\bc_{i}={-\beta\bt}/{(1-\beta)}$ and that of the outer ring $\bc_{i+1}={\beta\mathbf{b}}/{(1+\beta)}$. Note that the same launch trajectory is now a tangent to the inner ring with center as specified. Further $\ell\left( \bc_{i},\bd\right) =\beta$ and $\ell\left(\bc_{i+1},\bd\right) =\alpha\beta$, which satisfy their confidence region guarantees.

}

\bibliographystyle{IEEEtran}

\bibliography{references}

\begin{thebibliography}{10}
\providecommand{\url}[1]{#1}
\csname url@samestyle\endcsname
\providecommand{\newblock}{\relax}
\providecommand{\bibinfo}[2]{#2}
\providecommand{\BIBentrySTDinterwordspacing}{\spaceskip=0pt\relax}
\providecommand{\BIBentryALTinterwordstretchfactor}{4}
\providecommand{\BIBentryALTinterwordspacing}{\spaceskip=\fontdimen2\font plus
\BIBentryALTinterwordstretchfactor\fontdimen3\font minus
  \fontdimen4\font\relax}
\providecommand{\BIBforeignlanguage}[2]{{%
\expandafter\ifx\csname l@#1\endcsname\relax
\typeout{** WARNING: IEEEtran.bst: No hyphenation pattern has been}%
\typeout{** loaded for the language `#1'. Using the pattern for}%
\typeout{** the default language instead.}%
\else
\language=\csname l@#1\endcsname
\fi
#2}}
\providecommand{\BIBdecl}{\relax}
\BIBdecl

\bibitem{isit_12}
D.~Ramasamy and U.~Madhow, ``Can geographic routing scale when nodes are
  mobile?'' in \emph{Information Theory Proceedings (ISIT), 2012 IEEE
  International Symposium on}, July 2012.

\bibitem{Karp:GPSR}
B.~Karp and H.~T. Kung, ``{GPSR}: greedy perimeter stateless routing for
  wireless networks,'' in \emph{Proceedings of the 6th annual international
  conference on Mobile computing and networking}, 2000.

\bibitem{GuptaKumarWireless00}
P.~Gupta and P.~Kumar, ``The capacity of wireless networks,'' \emph{Information
  Theory, IEEE Transactions on}, mar 2000.

\bibitem{GuptaAsymptoticConn98}
------, ``Critical power for asymptotic connectivity,'' in \emph{Decision and
  Control, 1998. Proceedings of the 37th IEEE Conference on}, 1998.

\bibitem{Basagni:DREAM1998}
S.~Basagni, I.~Chlamtac, V.~R. Syrotiuk, and B.~A. Woodward, ``A distance
  routing effect algorithm for mobility ({DREAM}),'' in \emph{Proceedings of
  the 4th annual ACM/IEEE international conference on Mobile computing and
  networking}, 1998.

\bibitem{HazyStateLinkStateRoutingMobiHoc01}
C.~A. Santiv\'{a}\~{n}ez, R.~Ramanathan, and I.~Stavrakakis, ``Making
  link-state routing scale for ad hoc networks,'' in \emph{Proceedings of the
  2nd ACM international symposium on Mobile ad hoc networking \& computing},
  2001.

\bibitem{Li:ScalableLocationService:Mobicom:00}
J.~Li, J.~Jannotti, D.~S.~J. De~Couto, D.~R. Karger, and R.~Morris, ``A
  scalable location service for geographic ad hoc routing,'' in
  \emph{Proceedings of the 6th annual international conference on Mobile
  computing and networking}, 2000.

\bibitem{Flury:2006:MLS:EfficientAdHoc}
R.~Flury and R.~Wattenhofer, ``{MLS}: an efficient location service for mobile
  ad hoc networks,'' in \emph{Proceedings of the 7th ACM international
  symposium on Mobile ad hoc networking and computing}, 2006.

\bibitem{LLS:Abraham:2004}
I.~Abraham, D.~Dolev, and D.~Malkhi, ``{LLS}: a locality aware location service
  for mobile ad hoc networks,'' in \emph{Proceedings of the 2004 joint workshop
  on Foundations of mobile computing}, ser. DIALM-POMC '04, 2004.

\bibitem{AsymptoticCriticalGreedyGeo:TC}
P.-J. Wan, C.-W. Yi, L.~Wang, F.~Yao, and X.~Jia, ``Asymptotic critical
  transmission radii for greedy forward routing in wireless ad hoc networks,''
  \emph{Communications, IEEE Transactions on}, May 2009.

\bibitem{MobilityIncreasesCapacity:Grossglauser:2002}
M.~Grossglauser and D.~N.~C. Tse, ``Mobility increases the capacity of ad hoc
  wireless networks,'' \emph{IEEE/ACM Transactions on Networking}, August 2002.

\bibitem{Grossglauser:2006:EASE:EncounterHistory}
M.~Grossglauser and M.~Vetterli, ``Locating mobile nodes with {EASE}: learning
  efficient routes from encounter histories alone,'' \emph{IEEE/ACM
  Transactions on Networking}, June 2006.

\bibitem{Grossglauser:2003:AgeMatters:FRESH}
H.~Dubois-Ferriere, M.~Grossglauser, and M.~Vetterli, ``Age matters: efficient
  route discovery in mobile ad hoc networks using encounter ages,'' in
  \emph{Proceedings of the 4th ACM international symposium on Mobile ad hoc
  networking \& computing}, 2003.

\bibitem{Random:Waypoint:96}
D.~Johnson and D.~Maltz, ``Dynamic source routing in ad hoc wireless
  networks,'' in \emph{Mobile Computing}, T.~Imielinski and H.~Korth, Eds.,
  1996.

\end{thebibliography}

\end{document}